\documentclass[usenatbib]{mnras}
\usepackage{graphicx}
\usepackage{float}
\usepackage{amsmath}
\usepackage{amssymb}
\usepackage{textcomp}
\usepackage{multirow}  
\usepackage[caption=false]{subfig}
\usepackage{color}
\usepackage{gensymb}   
\usepackage{hyperref} 
\usepackage{xspace} 
\usepackage{afterpage} 

\newcommand{\comment}[1]{\ignorespaces}
\newcommand{\HI}{\textsc{Hi}\xspace}
\renewcommand{\sin}[1]{\text{Sin}\left(#1\right)\xspace}
\renewcommand{\cos}[1]{\text{Cos}\left(#1\right)\xspace}
\newcommand{\erf}[1]{\text{Erf}\left(#1\right)\xspace}
\newcommand{\heaviside}[1]{\mathbf{\Theta}\left(#1\right)\xspace}
\makeatletter
\newcommand\footnoteref[1]{\protected@xdef\@thefnmark{\ref{#1}}\@footnotemark}
\makeatother

\title[Outflows Quench Satellites of Low-Mass Hosts]{The Dual Role of Outflows in Quenching Satellites of Low-Mass Hosts: NGC 3109}

\author[C. T. Garling et al.]{Christopher T. Garling$^{1}$\thanks{E-mail: txa5ge@virginia.edu},
Annika H. G. Peter$^2$,
Kristine Spekkens$^{3,4}$,
David J. Sand$^5$,
Jonathan Hargis$^6$, \newauthor
Denija Crnojevi\'c$^7$,
Jeffrey L. Carlin$^8$ 
\\
$^1$Department of Astronomy, University of Virginia, 530 McCormick Road, Charlottesville, VA, 22904, USA \\
$^2$CCAPP, Department of Physics, and Department of Astronomy, The Ohio State University, 191 W. Woodruff Ave., Columbus, OH 43210, USA \\
$^3$Department of Physics and Space Science, Royal Military College of Canada P.O. Box 17000, Station Forces Kingston, ON K7K 7B4, Canada \\
$^4$Department of Physics, Engineering Physics and Astronomy, Queen’s University, Kingston, ON K7L 3N6, Canada \\
$^5$Department of Astronomy and Steward Observatory, University of Arizona, 933 N. Cherry Avenue, Tucson, AZ 85719, USA \\
$^6$Space Telescope Science Institute, 3700 San Martin Drive, Baltimore, MD 21218, USA\\
$^7$Department of Physics, University of Tampa, 401 West Kennedy Boulevard, Tampa, FL 33606, USA \\
$^8$Rubin Observatory Project Office, 950 N. Cherry Avenue, Tucson, AZ 85719, USA \\
}

\date{Accepted XXX. Received YYY; in original form ZZZ}

\pubyear{2022}

\begin{document}
\label{firstpage}
\pagerange{\pageref{firstpage}--\pageref{lastpage}}
\maketitle

\begin{abstract}
While dwarf galaxies observed in the field are overwhelmingly star-forming, dwarf galaxies in environments as dense or denser than the Milky Way are overwhelmingly quenched. In this paper, we explore quenching in the lower density environment of the Small-Magellanic-Cloud-mass galaxy NGC 3109 (M$_* \sim 10^8 \, \text{M}_\odot$), which hosts two known dwarf satellite galaxies (Antlia and Antlia B), both of which are \HI deficient compared to similar galaxies in the field and have recently stopped forming stars. Using a new semi-analytic model in concert with the measured star formation histories and gas masses of the two dwarf satellite galaxies, we show that they could not have been quenched solely by direct ram pressure stripping of their interstellar media, as is common in denser environments. Instead, we find that separation of the satellites from pristine gas inflows, coupled with stellar-feedback-driven outflows from the satellites (jointly referred to as the starvation quenching model), can quench the satellites on timescales consistent with their likely infall times into NGC 3109's halo. It is currently believed that starvation is caused by ``weak" ram pressure that prevents low-density, weakly-bound gas from being accreted onto the dwarf satellite, but cannot directly remove the denser interstellar medium. This suggests that star-formation-driven outflows serve two purposes in quenching satellites in low-mass environments: outflows from the host form a low-density circumgalactic medium that cannot directly strip the interstellar media from its satellites, but is sufficient to remove loosely-bound gaseous outflows from the dwarf satellites driven by their own star formation.  
\end{abstract}


\section{Introduction}
One of the most important questions in galaxy evolution is why galaxies transition from blue and actively star-forming to red and quiescent. While quenched high-mass (stellar mass M$_* \geq 10^{11}$ M$_\odot$) galaxies are prevalent in the field, suggesting that they may be quenched by internal mechanisms like feedback from active galactic nuclei and virial shock heating \citep[e.g.,][]{Birnboim2003,DiMatteo2005,Wang2008,Martig2009,Keres2009,Dave2017,Schreiber2018}, galaxies with lower masses are very rarely observed to be quenched when in isolation \citep{Haines2007,Geha2012,Kawinwanichakij2017}. Using catalogs of multi-band galaxy photometry derived from wide-field optical imaging surveys, it has been shown that the galactic environment (and in particular, host halo mass) is the primary driver of quenching in these lower-mass galaxies \citep[e.g.,][]{Woo2013,Zu2016,Kawinwanichakij2017,Moutard2018,Papovich2018,Davies2019}. This conclusion is consistent with the long-standing observation that red, quenched galaxies are abundant in dense cluster environments \citep[e.g.,][]{Dressler1980,Butcher1984,Hogg2003,Kauffmann2004,Cooper2006,Haines2007}. \par

Much of the recent progress made in understanding environmental star formation quenching is due to new wide-area surveys which provide data for large statistical samples of galaxies, typically reaching stellar masses of $10^8$ M$_\odot \leq$ M$_* \leq 10^{10}$ M$_\odot$ at the faint end \citep[e.g.,][]{Driver2011,Taylor2011,Chambers2016,Kuijken2019,Zou2019}. Below these masses, the study of star formation quenching has mostly been limited to the Local Volume, with particular attention paid to satellites of the Milky Way (MW) and Andromeda (M31). Due to the close proximity of these dwarf satellites, a wide range of observational and theoretical tools can be leveraged to learn about their quenching processes \citep[see, e.g.,][]{Mayer2006,Grcevich2009,Nichols2011,Rocha2012,Gatto2013,Slater2014,Weisz2015,Wetzel2015,Fillingham2015,Fillingham2016,Fillingham2019,Buck2019c,Digby2019,Garrison-Kimmel2019,Akins2021,Applebaum2021}. These studies indicate that there is a divide in the quenching pathways of the Local Group dwarf satellites around stellar masses of $10^5$ M$_\odot$, where the least massive (``ultra-faint") dwarf satellites were quenched by reionization \citep[e.g.,][]{Bullock2000,Benson2002,Brown2014,Weisz2014b,Weisz2015,RodriguezWimberly2019,Sand2022}, while the more massive (``classical") dwarf satellites were quenched later via environmental processes. This later environmental quenching is typically attributed to ram pressure stripping (RPS; \citealt{Gunn1972,Larson1980}) caused by the host's hot gas halo \citep[][]{Gatto2013,Slater2014,Emerick2016,Fillingham2016}. In particular, the intermediate mass ($10^5$ M$_\odot \leq$ M$_* \leq 10^8$ M$_\odot$) dwarf satellites of the MW exhibit short ($\sim2$ Gyr) quenching timescales that are best explained by direct stripping of the interstellar media of the dwarfs via RPS or tidal stripping \citep[e.g.,][]{Wetzel2015,Fillingham2015,Fillingham2019,Baxter2021}. This sort of ``strong" RPS also appears in observations and simulations of dense cluster environments, where RPS is highly effective due to the presence of a hot intracluster medium \citep[e.g.,][]{Lotz2019,Roberts2019,Tremmel2019,Tonnesen2019}.\par

However, the satellites of the MW seem to be somewhat unique in this regard \citep{Geha2017,Mao2021,Karunakaran2021,Carlsten2021b,Carlsten2022,Samuel2022}. By comparing star-forming and quenched populations of bright (M$_* \geq 10^8$ M$_\odot$) satellite galaxies measured in wide-area surveys, it has been found that massive dwarfs have longer quenching timescales \citep[$\sim$4--6 Gyr;][]{Wetzel2012,Wheeler2014}. These longer quenching timescales are consistent with starvation, wherein accretion of gas is ceased after infall to the host but the satellite continues to form stars until its gas reservoir is depleted by star-formation-driven outflows. Quenching via starvation is theorised to require ``weak" environmental RPS, wherein low-density gas at large radii from the dwarf can be removed but the denser interstellar gas of the dwarf cannot \citep[e.g.,][]{Maier2019}, such that star formation can continue after infall but gas loosened by stellar feedback (e.g., supernovae) is lost. Thus both ``strong" RPS and starvation require the presence of a circumgalactic medium around the host, given our current theoretical understanding of these quenching mechanisms. Such a circumgalactic medium could be generated by stellar-feedback-driven gas outflows from the host.\par

There is an additional factor to be considered when studying environmental quenching that we have so far only mentioned in passing; the dependence of the quenching timescales and mechanisms on environmental density. Galaxy clusters inhabit the densest environments, while an isolated low-mass galaxy with a few dwarf satellites is a low-density environment. As previously mentioned, quenched dwarf satellites are prevalent in dense environments, while equal-mass dwarfs are predominantly star-forming in the field (i.e., when they are far from more massive galaxies; \citealt{Geha2012}). However, it is unclear how environmental quenching scales to lower host masses.\par 

Such low-density environments are poorly represented in the literature, as the hosts and satellites are intrinsically faint, severely limiting the distances to which such systems can be studied. However, several pioneering surveys have begun to extend the study of environmental quenching to low-density environments (e.g., MADCASH and LBT-SONG; \citealt{Carlin2016,Carlin2019,Carlin2021,Garling2020,Davis2021}), with initial results indicating that such low-mass hosts can, indeed, quench their satellites, with starvation being the most likely quenching mechanism. The theoretical literature considering environmental quenching in low-density environments is similarly sparse, but recent simulations show that these low-mass hosts may be able to sustain hot halos of circumgalactic gas, which is generally thought to be a requirement for these types of environmental quenching processes \citep{Jahn2019,Jahn2021}. Clearly our picture of star formation quenching is incomplete, in particular at low satellite and host masses. Additionally, the studies that probe low satellite masses are generally confined to satellites of the MW and M31 and are therefore incomplete in terms of galactic environment; this is a particular problem given that host halo mass is the primary driver of satellite quenching, as mentioned above. As a result, there is much to be gained by extending the study of low-mass (M$_* \leq 10^8$ M$_\odot$) dwarf satellite quenching beyond the MW. \par

In order to assess the efficacy of different quenching mechanisms in low-density environments, we study the NGC~3109 dwarf association, which is ideal for this kind of study. NGC 3109 itself is similar in stellar mass to the Small Magellanic Cloud (SMC) but has not yet been accreted by a larger galaxy, making this a very low-density environment. NGC 3109 is known to host two classical dwarf analogs, the Antlia and Antlia B dwarf satellite galaxies (see \S\ref{section:obsprop} for more discussion of the observational properties of the system). NGC 3109 can therefore be considered a dwarf group or association \citep[e.g.,][]{Tully2006,Stierwalt2015,Pearson2016}.\par

The Antlia and Antlia B dwarf satellite galaxies both have low present-day \HI masses relative to the typical values for isolated dwarfs of similar stellar mass (e.g., \citealt{Papastergis2012,Bradford2015,Scoville2017}), suggesting that they have been affected by environmental quenching. Based on the M$_*$ -- M$_\HI$ relation of \cite{Bradford2015}, Antlia has only $2\%$ of the \HI mass of field dwarfs with comparable stellar mass, while Antlia B has roughly $15\%$ of the \HI mass of comparable field dwarfs. While the M$_*$ -- M$_\HI$ relation is uncertain at the low-mass end due to the limited sample of low-mass field dwarfs, it is clear that Antlia and Antlia B are significant outliers from the field population, prompting us to examine how their \HI could have been depleted. \par

As the system is nearby (with distance $\sim1.3$ Mpc; \citealt{Dalcanton2009}) and well-studied observationally (with measured star formation histories from resolved stellar populations, stellar masses, line-of-sight velocities, etc.), we have excellent data with which to set up a theoretical experiment. These observational data allow us to select analog dwarf galaxy systems from a cosmological simulation, which we use in concert with a simple semi-analytic model to study the gas mass evolution of the satellites after infall. Through this semi-analytic model we can assess which quenching mechanisms are most important in this system, providing a template that can be applied to other systems to search for a model which is generally successful.  

\begin{figure*} 
  \centering
  \includegraphics[width=\textwidth,page=1]{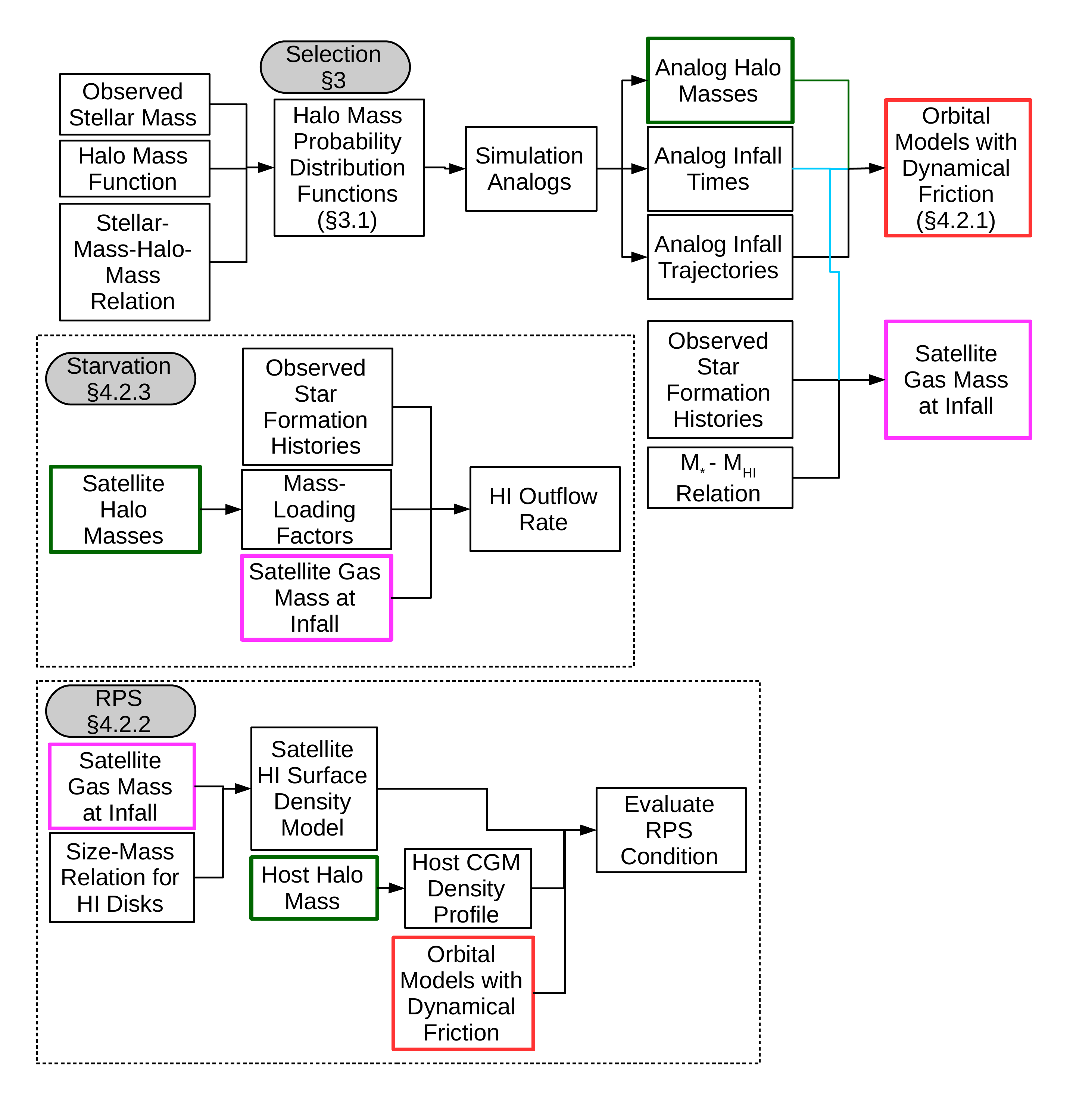}
  \caption{Flow chart summarizing the relationships between different components of our semi-analytic quenching model. We begin at the top of the figure with our selection procedure discussed in \S\ref{section:illustris}. Colored boxes indicate intrinsic (e.g., halo masses) or derived (e.g., satellite gas masses at infall) properties of individual analog systems from the IllustrisTNG simulation that are key inputs to our starvation and RPS quenching models. We then run our semi-analytic quenching model on the analog systems from the simulation that pass our selection criteria. The relationship between the inputs and outputs for the quenching models are shown in their own labelled flow charts.}
  \label{figure:flowchart}
\end{figure*}

\subsection{This approach}
We adopt a semi-analytic approach to study satellite quenching in the NGC 3109 system. The primary cosmological ingredient to our analysis is the merger history of NGC 3109; of particular importance is exactly \emph{when} Antlia and Antlia B fell into NGC 3109, because these infall times define the timescales over which our environmental quenching models can act. For this purpose, we draw analogs of NGC 3109 and its satellites from big-box hydrodynamic simulations. These simulations are sufficient to resolve the satellites in dark matter, but they generally have poorly-resolved baryonic components. As such, it is necessary to post-process the baryonic components of the satellite galaxies, which we accomplish using simple analytic models coupled to the observed star formation histories (SFHs). These theoretical tools, when combined with the observed properties of the NGC 3109 system (e.g., phase-space coordinates, SFHs, and present-day \HI gas masses), allow us to discern the most likely quenching pathway for the satellites.\par

A schematic illustrating how our model operates is shown in Figure \ref{figure:flowchart}. It begins with identifying analogs of the NGC 3109 system that host \emph{either} an Antlia or Antlia B analog satellite at present day in the simulation. We do not require the system to host \emph{both} an Antlia and an Antlia B analog; justification for this choice is given in \S\ref{subsubsection:jointsample}. This process relies on the measured stellar masses for each galaxy, which are converted to halo mass probability density functions through a halo mass function and an empirical relation between galaxy stellar mass and halo mass; analogs are then selected from the simulation on the basis of halo mass. The process of analog selection is explained in \S \ref{section:illustris}. \par

From these simulated analogs we establish the initial conditions for our semi-analytic models by incorporating other empirical and theoretical relations (\S\ref{subsection:ic}). The stellar mass of the analog satellites at infall is set by integrating the observational SFHs for the real satellites up until infall. We then use the empirical relation between stellar mass and \HI mass for isolated dwarfs to set the \HI mass of the satellites at infall, which is used for both the RPS (\S\ref{subsubsection:RPS}) and starvation (\S\ref{subsubsection:starvation}) calculations. For starvation, the only other important variable quantity is the effective mass-loading factor, which is the constant of proportionality between the star formation rate and the star-formation-driven \HI mass outflow rate (i.e., $\eta$ such that $d\text{M}_\HI/dt \propto -\eta \, d\text{M}_*/dt$). This mass-loading factor can generally be tied to the halo mass of the satellite. Coupling the initial satellite \HI mass, mass-loading factor, observed SFH, and infall time gives us enough information to calculate the gas mass-loss rate due to starvation. \par

For RPS, we require a few more components. As RPS is dependent on the satellite orbits, and the time resolution of the simulation output is poor, we resimulate the orbits of the satellite analogs with dynamical friction as discussed in \S \ref{subsubsection:dynamicalfriction}. RPS is also dependent on the satellite gas surface density profile, which we couple to the initial \HI mass following relations from the literature. The density profile of the host circumgalactic medium is also important, for which we adopt a literature relation tied to the host halo mass. Given the satellite orbit and the gas profiles of the host and satellite, we can determine how much gas is removed by RPS after infall of the analog satellites. Due to our semi-analytic method, we are able to separate the effects of starvation and RPS to determine which is more effective. \par

In \S\ref{section:results}, we present results from the fiducial model and explore tidal stripping, alternate quenching model parameters, and uncertainties in the quenching models. Our model clearly indicates that starvation is more effective at removing gas from satellites of low-mass hosts than RPS. We further show that our fiducial RPS model, in the absence of starvation, produces Antlia and Antlia B analogs that are too gas-rich across the entire range of infall time of our simulated analogs, indicating the importance of starvation for quenching such satellites of low-mass hosts. We comment on the implications of this result in \S\ref{section:conclusion}. \par

\section{Observational Properties} \label{section:obsprop}
One of the principle reasons for using the NGC 3109 system as a case-study for quenching in low-mass systems is the abundance of data available for the host and satellites, including stellar masses, star formation histories, and \HI masses. In this section, we highlight the observational properties that will be relevant for our theoretical work. These quantities are also presented in Table \ref{Table:systemproperties}. \par

The primary observable we use to select analogs from the simulations are the observed stellar masses for NGC 3109 and its satellites (see \S\ref{section:illustris} for a discussion of our selection procedure). We adopt a stellar mass of $7.3\pm0.9 \, \times 10^6$ M$_\odot$ for the Antlia dwarf satellite based on its resolved SFH \citep{McQuinn2010,McQuinn2010a}. Thus Antlia sits at the upper end of the classical dwarf regime, with a stellar mass within an order of magnitude of the Leo I, Fornax, and Sculptor dwarf satellites of the MW \citep[][and references therein]{McConnachie2012}. For the Antlia B dwarf satellite, we adopt $6^{+4}_{-3} \, \times 10^5$ M$_\odot$ for its stellar mass, based on the aperture photometry of \cite{Sand2015a}. This work assumed a stellar mass-to-light ratio of $\Upsilon=1$. This also makes Antlia B a classical dwarf analog, with a stellar mass comparable to the Draco, Ursa Minor, Sextans I, Carina, Canes Venatici I, and Leo II dwarf satellites of the MW \citep{McConnachie2012}. \par

It is worth noting that the stellar mass of NGC 3109 is significantly uncertain; \cite{McConnachie2012} lists $7.6\times10^7$~M$_{\odot}$ and cites \cite{BlaisOuellette2001}, who fit the stellar mass-to-light ratio of NGC 3109 to their \HI radial velocity profile under several different models for the halo density profile. However, these stellar-mass-to-light ratios have an order of magnitude spread depending on which halo profile is considered, and several of these profiles have similar goodness of fit. We derive the stellar mass of NGC 3109 by adopting photometry from \cite{Cook2014}, which is more precise than the photometry used in \cite{BlaisOuellette2001}, and estimate the stellar-mass-to-light ratio using the color-dependent relationships in \cite{GarciaBenito2019}.  Using this method, we find a stellar mass-to-light ratio of $\Upsilon=0.8$, which implies a stellar mass of $1.4\times10^8$ M$_{\odot}$. Adopting the lognormal spread of 0.11 dex in the color-stellar-mass-to-light ratio from \cite{GarciaBenito2019}, the $1-\sigma$ range of stellar masses is $1.1\times10^8$ to $1.8\times10^8$ M$_{\odot}$. These estimates are roughly twice the value of $7.6\times10^7$ M$_{\odot}$ from \cite{McConnachie2012} and \cite{BlaisOuellette2001}, but are based on better photometry \citep{Cook2014} and stellar mass-to-light ratios \citep{GarciaBenito2019}. \par 

We can additionally compare to the $K_s$ band luminosity, which has a fairly constant stellar-mass-to-light ratio of 0.6 with $\sim0.1$ dex scatter \citep{McGaugh2014}. Adopting the $K_s$ mag from the 2MASS Large Galaxy Atlas \citep{Jarrett2003} we find a stellar mass of $2.4\times10^8$ M$_*$, with a $1\sigma$ range of $1.9\times10^8$ to $3.0\times10^8$ M$_*$. This is higher than the estimate from the optical photometry, but they agree at a $1.2\sigma$ level and both prefer a higher stellar mass than that given in \cite{McConnachie2012} and \cite{BlaisOuellette2001}. Both the optical \citep{Cook2014} and infrared \citep{Jarrett2003} magnitudes were corrected for Galactic extinction using the \cite{Cardelli1989} extinction law and the \cite{Schlegel1998} dust maps with the updated scaling from \cite{Schlafly2011}. \par 

Our revised stellar mass estimates from the optical \citep{Cook2014} and infrared \citep{Jarrett2003} photometry are sufficiently consistent that it makes no significant difference to our conclusions in \S\ref{section:results} which we use. We adopt the stellar mass estimate for NGC 3109 based on the \cite{Cook2014} magnitudes and \cite{GarciaBenito2019} stellar mass-to-light ratio for our analysis. This stellar mass is comparable to that of the Small Magellanic Cloud \citep[SMC;][]{McConnachie2012}. We note that RPS is more effective for larger host stellar masses, so by assuming a larger stellar mass for NGC 3109 we are making RPS more effective than it would be if we adopted the \cite{BlaisOuellette2001} stellar mass.  \par

While the present-day stellar masses of the satellite galaxies are important for selecting analogs from the cosmological simulation (\S\ref{section:illustris}), how the satellites built up their stellar masses over time (i.e., their SFHs) matters for our implementation of quenching via starvation (\S\ref{subsubsection:starvation}). Normally in fully semi-analytic models, the SFH of galaxies is self-consistently evolved depending on other variables like neutral or molecular hydrogen masses. For a case-study like this, such an approach is undesirable because the real satellites have a fixed intrinsic SFH which may not be well-sampled by the simulated analogs. Fortunately, both Antlia and Antlia B have measured SFHs based on resolved stars. We therefore adopt these SFHs directly (from \citealt{Weisz2011} for Antlia and \citealt{Hargis2020} for Antlia B) so that the SFH is not a free parameter in our model. These SFHs are utilized in our model of starvation, wherein stellar feedback from young stars (primarily in the form of supernovae) ejects gas from the satellite, slowly quenching its star formation after infall as (under the model assumptions) the satellite cannot accrete more gas from its environment while inside the halo of NGC 3109. The SFHs of both Antlia and Antlia B show little star formation in the last few Gyr, suggesting that some environmental quenching process is responsible. \par

The measured \HI masses of Antlia and Antlia B are also suggestive of environmental effects, as they are lower than those of isolated dwarfs in the field of similar stellar mass \citep[e.g.,][]{Papastergis2012,Bradford2015,Scoville2017}. Typically such isolated dwarfs have at least twice as much mass in \HI as in stars, but this is not the case for the dwarf satellites of NGC 3109. Antila has a measured \HI mass of $6.8 \pm 1.4 \, \times 10^5$ M$_\odot$ \citep{Barnes2001}, roughly 10\% of its stellar mass, while Antlia B has an \HI mass of $2.8 \pm 0.2 \, \times 10^5$ M$_\odot$ \citep{Sand2015a}, which is roughly half of its stellar mass. \par

With the measured \HI masses, we also get line-of-sight velocities for the dwarfs \citep{Barnes2001,Ott2012,Sand2015a}, which we use in concert with the 2D separation in the plane of the sky between NGC 3109 and its satellites to further constrain our simulated samples (see \S\ref{subsection:analogselection} and Appendix \ref{appendix:pq}). While NGC 3109, Antlia, and Antlia B all have robust distance measurements that indicate they are associated \citep{Dalcanton2009,Hargis2020}, the uncertainties are still large enough that there is little to be gained (in a statistical sense) from including the distances of the galaxies into these constraints. \par

\section{Simulations} \label{section:illustris}

\begin{table*}
  \caption{Assumed properties of the host, NGC 3109, and the satellites, Antlia and Antlia B. References: (1) \citealt{Dalcanton2009} (2) \citealt{Cook2014,GarciaBenito2019} (3) \citealt{Barnes2001} (4) \citealt{Ott2012}  (5) \citealt{McQuinn2010a} (6) \citealt{Hargis2020} (7) \citealt{Sand2015a}}
  \centering
    \hspace*{-4em}
    \begin{tabular}{c c c c c c c c}
      \hline
      \hline
      Object & Distance & R.A. & Decl. & M$_*$ & M$_{\HI}$ & V$_{\text{LOS}}$ & References \\
       & [Mpc] & [hms] & [dms] & [$10^5$ M$_{\odot}$] & [$10^5$ M$_{\odot}$] & [km/s] \\
      \hline
      NGC 3109 & 1.29 $\pm0.02$ & $10^{\text{h}}03^{\text{m}}07^{\text{s}}$ & $-26\degr09\arcmin36\arcsec$ & 1400$^{+400}_{-300}$ & $3800\pm500$ & $405\pm2$ & (1) (1) (1) (2) (3) (4)\\ 
      Antlia & 1.29 $\pm0.02$ & $10^{\text{h}}04^{\text{m}}04^{\text{s}}$ & $-27\degr19\arcmin55\arcsec$ & 73 $\pm 9$ & $6.8\pm1.4$ & $363\pm2$ & (1) (1) (1) (5) (3) (4) \\ 
      Antlia B & $1.35\pm0.06$ & $09^{\text{h}}48^{\text{m}}56^{\text{s}}$ & $-25\degr59\arcmin24\arcsec$ & 6$^{+4}_{-3}$ & $2.8\pm0.2$ & $376\pm2$ & (6) (7) (7) (7) (7) (7)  \\
      \hline
    \end{tabular}
  \label{Table:systemproperties}
\end{table*}

We use simulations to select an ensemble of NGC 3109-like analogs, the key input for our semi-analytic quenching exploration.  Simulations allow us to sample the range of satellite infall times and orbits that are consistent with the observed properties of the satellites (e.g., their angular separation and relative line-of-sight velocity from the host). Infall times matter because they set the clock for the timescale of quenching, and the orbits matter because RPS is orbit-dependent. Applying the same semi-analytic model across the sample of analog systems yields a statistical exploration of quenching mechanisms and timescales.  With the probabilistic analog selection procedure we describe below, we can more accurately examine the probability distributions of quantities like the satellite infall times and the \HI mass loss due to RPS. \par

We utilize the public IllustrisTNG cosmological simulations to select analogs of the NGC 3109 system \citep{Nelson2018,Naiman2018,Springel2018,Pillepich2018a,Marinacci2018}. IllustrisTNG is simulated with a $\Lambda$CDM cosmology with parameters from the Planck 2015 results \citep[$h$=0.6774;][]{Planck2016}; we will adopt this cosmology throughout. IllustrisTNG includes hydrodynamics, with a fiducial physics model presented in \cite{Weinberger2017} and \cite{Pillepich2018}. We use the TNG100 run, which simulates a comoving box of volume 110.7 Mpc\textsuperscript{3} with $1820^3$ particles each for dark matter and gas and twice as many tracers that are used to track the Lagrangian evolution of the gas \citep{Genel2013}. This simulation suite is well-matched to our science goals, because it has sufficiently high resolution in dark matter over a sufficiently large volume for us to obtain a statistically significant set of NGC 3109 system analogs. \par

We utilize the friends-of-friends (FoF) group catalogs to identify isolated host systems, the \textsc{subfind} subhalo catalogs to extract subhalo properties, and the \textsc{sublink} merger trees \citep{Rodriguez-Gomez2015} to track the evolution of the subhalos through time. The friends-of-friends halo finder requires a minimum of 32 particles, corresponding to $2.4 \times 10^8 \ \text{M}_{\odot}$ if the particles are all dark matter, while \textsc{subfind} requires a minimum of 20 particles that are gravitationally bound, corresponding to a halo mass lower bound of $1.5 \times 10^8 \ \text{M}_{\odot}$. In \S \ref{subsubsection:hmp} we show that the expected halo mass for Antlia B, the least massive galaxy in the NGC 3109 system, is a factor of 20 larger than this, giving us confidence that the dark matter halos of satellites like Antlia and Antlia B in the simulation will be sufficiently resolved for our purposes (see Figure \ref{figure:flowchart} for an overview of how we use the halo quantities from the simulation). \par

To define a set of analogs, the principal observables we have for NGC 3109 and its satellites are their stellar masses (see Table \ref{Table:systemproperties}). The easiest way to identify analogous systems in TNG100 would be to find systems with similar stellar masses in the simulation catalogs. However, the baryonic particle masses in TNG100 are $1.4\times10^6$ M$_{\odot}$, so that the stellar masses of Antlia and Antlia B analogs will be poorly resolved. Instead, we can infer the halo masses of NGC 3109 and its satellites from their stellar masses given a stellar-mass-halo-mass (SMHM) relation, a halo mass function (HMF), and a subhalo mass function (SMF). The conversion of observed stellar mass to approximate halo mass must be considered carefully. As noted in \cite{Dooley2017b}, \cite{Somerville2018}, \cite{Jethwa2018}, and other works, the SMHM relation cannot simply be inverted to obtain a halo mass from an observed stellar mass because the relationship has intrinsic scatter. This leads to significant Eddington bias, as there are many more low-mass galaxies to be up-scattered than high-mass galaxies to be down-scattered, indicating that a naive inversion of the SMHM relation would overestimate the typical halo mass of galaxies observed at a fixed $\text{M}_*$. This effect can be mitigated by including models for the HMF and SMF as shown below.

\subsection{Selection of Analogs via Halo Mass Probabilities} \label{subsubsection:hmp}

The probability density function (PDF) of a halo mass given a stellar mass, $P(\text{M}_h | \text{M}_*)$ can be related to the PDF of a stellar mass given a halo mass from the SMHM relation, $P(\text{M}_* | \text{M}_h)$, and the halo mass function, $d\text{N}/d\text{M}_h \propto P(\text{M}_h)$, through Bayes' theorem as $P(\text{M}_h | \text{M}_*) \propto P(\text{M}_* | \text{M}_h) \ P(\text{M}_h)$. We additionally add a reionization quenching model that describes the fraction of halos of mass M$_h$ that are luminous, denoted $f_{\text{lum}} (\text{M}_h)$. For a constant lognormal scatter in the SMHM relation in dex of $\sigma$ and a fiducial stellar mass of M$_*$, we can write the PDF of a stellar mass given a halo mass as

\begin{equation} \label{equation:stellarmassprob}
\begin{aligned}
    P(\text{M}_* | \text{M}_h) &= \frac{f_{\text{lum}}(\text{M}_h)}{\text{M}_* \, \sigma \ \text{ln}10 \, \sqrt{2 \pi}} \\
    & \times \text{exp} \left[ \frac{- \left(\text{log}_{10} \text{M}_* - \text{log}_{10} \text{SMHM}\left(\text{M}_h\right) \right)^2}{2 \sigma^2} \right] \\
\end{aligned}
\end{equation}

\noindent where SMHM(M$_h$) is the median stellar mass from the SMHM relation for a halo mass of M$_h$. In order to apply Bayes' theorem to find $P(\text{M}_h | \text{M}_*)$ we must calculate the Bayesian evidence to properly normalize the PDF, which can be written as

\begin{equation} \label{equation:bayesianevidence}
\begin{aligned}
        \text{A} &= \int_0^\infty \int_0^\infty P(\text{M}_* | \text{M}_h) \ \frac{d\text{N}}{d\text{M}_h} \, P(\text{M}_*) \, d\text{M}_h \, d\text{M}_*  \\
        \end{aligned}
\end{equation}

\noindent where $P(\text{M}_*)$ is the PDF for the stellar mass of the object in question. We can then write the conditional PDF of $\text{M}_h$ given $\text{M}_*$ as 

\begin{equation} \label{equation:halomassprob}
\begin{aligned}
        P(\text{M}_h|\text{M}_*) &= \frac{1}{\text{A}} \int_0^\infty P(\text{M}_*|\text{M}_h) \ \frac{d\text{N}}{d\text{M}_h}  \ P(\text{M}_*) \ d\text{M}_*
        \end{aligned}
\end{equation}

\noindent One useful application of this equation is to compute the expectation value for the halo mass of a galaxy given its stellar mass, which can be written as

\begin{equation} \label{equation:halomassexpectation}
\begin{aligned}
        \langle \text{M}_h \rangle &= \int_0^\infty \int_0^\infty \text{M}_h \, P(\text{M}_h|\text{M}_*) \, d\text{M}_h \, d\text{M}_* \\
        &= \frac{1}{\text{A}} \int_0^\infty \int_0^\infty \text{M}_h \ P(\text{M}_* | \text{M}_h) \ \frac{d\text{N}}{d\text{M}_h} \ P(\text{M}_*) \ d\text{M}_h \, d\text{M}_*
        \end{aligned}
\end{equation}

\noindent When considering the halo mass PDFs for subhalos, the SMF, $d\text{N}/d\text{M}_{h,\text{sat}}(\text{M}_{h,\text{host}})$, should be used instead of the HMF. We use the form of the halo mass function from \cite{Sheth2001} with the transfer function from \cite{Eisenstein1999} to generate the halo mass function, the SMF and $f_{\text{lum}} (\text{M}_h)$ from \cite{Dooley2017b}, and the \cite{Moster2013} SMHM relation. We assume a constant 0.2 dex scatter in stellar mass at fixed M$_h$ \citep[e.g.,][]{Behroozi2013,Dooley2017b}. There is evidence for increased scatter in the SMHM below M$_h \sim 10^{11.5} \ \text{M}_{\odot}$, but we choose to keep a constant scatter for easier comparison to other work. The expected M$_h$ derived from Equation \ref{equation:halomassexpectation} are typically $\sim10\%$ lower than the result from a simple inversion of the \cite{Moster2013} SMHM relation in the range of stellar mass considered here. \par

With PDFs for the halo masses of NGC 3109 and its satellites, we are able to select halo mass ranges for each based on percentiles of enclosed probabilities, e.g. $68\%$ corresponding to the 1-$\sigma$ range for a Gaussian distribution. It is typical when selecting analogs of observational systems from simulations to choose a narrow range around the expected halo mass and approximate all simulated analogs as being equally likely to represent the observed system. However, this approach both limits the sample size of simulated analogs and neglects the tails of the halo mass PDF. We instead derive analog halo mass selection ranges as the intervals that contain 99.7$\%$ of the probability from the full halo mass PDFs (equivalent to $\pm3$-$\sigma$ for a Gaussian distribution), and fully propagate the probability that each simulated system represents the observed system through our analysis. We present our expectation values for M$_h$ and selection ranges in Table \ref{Table:m_halo_constraints}, along with the results from a naive inversion of the SMHM relation for comparison.\par

\subsection{Halo Overdensity Definitions} \label{subsubsection:so}
There is an additional complication here related to the definition of the halo mass. This matters because we match the stellar and dark matter halo masses of satellites at the time of satellite infall, and because Antlia is a major merger event for NGC 3109. Often halo masses are defined relative to spherical overdensity (SO) criteria; e.g., M$_{200\text{c}}$ is defined to be the total halo mass enclosed in a sphere whose average density is 200 times the critical density of the Universe, with a corresponding radius R$_{200\text{c}}$. Such SO masses and radii are only calculated in IllustrisTNG for the FoF group catalogs, while the \textsc{subfind} subhalo catalogs and \textsc{sublink} merger trees contain no such quantities, having instead only total gravitationally-bound masses; these are not directly comparable to SO quantities. Generally, models of the type used in Equation \ref{equation:halomassprob} are expressed in terms of SO mass definitions, with M\textsubscript{200c} being the most common mass definition -- we adopt M\textsubscript{200c} to evaluate Equation \ref{equation:halomassprob}, as all of the component models support this definition. Given that the stellar masses of satellites are more closely tied to their halo masses at infall than at present day \citep[e.g.,][]{Reddick2013,RodriguezPuebla2017,Campbell2018,Behroozi2019,Buck2019c,Moster2021,Wang2021}, what we would like in order to perform our analog selections are present-day M\textsubscript{200c} values for the hosts, and M\textsubscript{200c} values at infall for the satellites. \par

\begin{figure}
  \centering
  \includegraphics[width=0.45\textwidth,page=1]{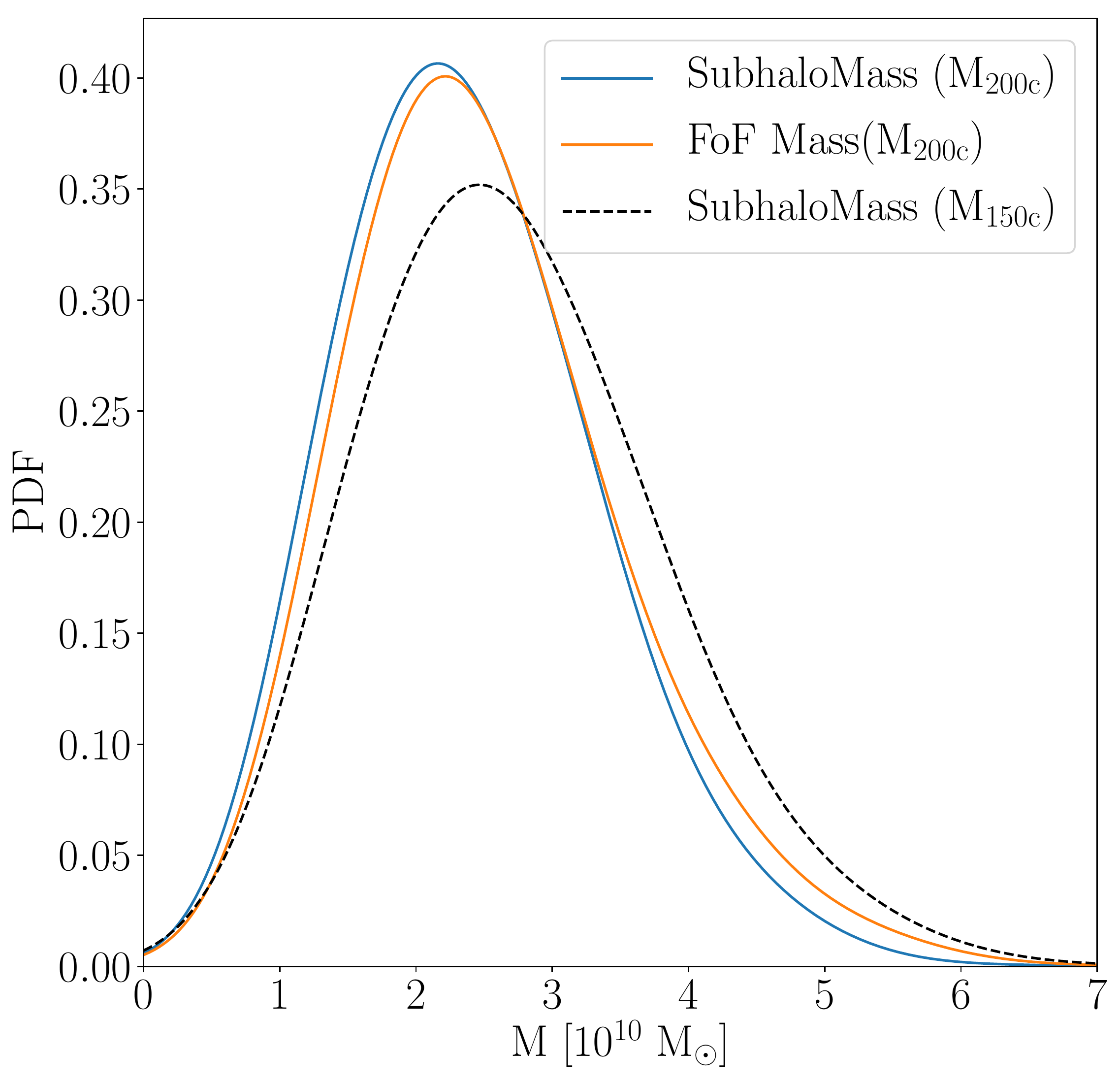}
  \caption{Comparison of the FoF M\textsubscript{200c} values (orange line) to the \textsc{sublink} subhalo masses (black dashed line) for our NGC 3109 analogs prior to the infall of Antlia or Antlia B. These distributions are weighted by $P(\text{M}_{h,\text{host},i}|\text{M}_{*,\text{host}})$ for each host $i$, as given in Equation \ref{equation:halomassprob}. If we assume that the \textsc{sublink} subhalo masses are approximately equal to M\textsubscript{150c}, we obtain the blue line when we convert them to M\textsubscript{200c}. This distribution is a good approximation of the FoF M\textsubscript{200c} distribution, and using the subhalo masses for host selection allows us to avoid including substructure in the host masses, as would be the case if we adopted the FoF masses.}
  \label{figure:m150c_conv}
\end{figure}

The obvious choice would be to use the M\textsubscript{200c} values from the FoF catalogs for the hosts at present-day, but these will include the masses of all subhalos. This is normally fine in the limit of M$_{host}>>$ M$_{sat}$, but given that Antlia is expected to have a halo mass $\sim25\%$ that of NGC 3109 (i.e., Antlia's infall constitutes a major merger), using these FoF masses for the hosts may bias our host selection as the FoF masses are correlated with the total mass in substructure. Instead, we identify the most massive subhalo in the \textsc{subfind} catalogs to be the central subhalo for a given satellite's FoF group. The subhalo mass for the central will consider particles that are only bound to the central and no other substructure, allowing us to avoid this bias. In order to convert the \textsc{subfind} subhalo mass to an SO mass, we approximate M\textsubscript{\textsc{subfind}} $\approx$ M\textsubscript{150c}. Prior to Antlia's infall, we find good agreement between the FoF M\textsubscript{200c} values and those obtained by this approximation, where we convert M\textsubscript{150c} to M\textsubscript{200c} assuming the concentration-mass relation of \cite{Diemer2019a}. We show the probability-weighted halo mass distribution for our NGC 3109 analogs prior to the infall of Antlia or Antlia B in Figure \ref{figure:m150c_conv}, along with the distribution derived for M\textsubscript{200c} assuming that the subhalo masses are approximately M\textsubscript{150c}. For the range of halo masses (and thus, concentrations) considered here,  M\textsubscript{200c}/M\textsubscript{150c} is typically 0.93 -- 0.96 at $z=0$, so this correction is modest and allows us to avoid biasing our host selection by including the mass of Antlia in the mass of NGC 3109. \par

\subsection{Summary of Analog Selection} \label{subsection:analogselection}
With these complexities explained, our analog selection proceeds as follows. We first identify present-day analogs of NGC 3109 by finding all central subhalos of the FoF groups and assuming their \textsc{subfind} masses are approximately equal to M\textsubscript{150c}. We convert these to M\textsubscript{200c} using the concentration-mass relation of \cite{Diemer2019a}, and save all the centrals that have halo masses within the 99.7\% credible interval for NGC 3109, given in Table \ref{Table:m_halo_constraints}. For every subhalo of these centrals identified at present-day, we identify each infall event, where the subhalo transitions from being its own central in the FoF catalogs to being a subhalo of the present-day host. We refer to the first infall as being the earliest such event. Note that by requiring subhalos to be associated with NGC 3109-like hosts at present-day, we are excluding splashback halos which may be beyond the host's virial radius at present-day but on a bound orbit. We are also excluding orphaned subhalos, which do not survive until present day. There are hints that Antlia could be tidally disrupting \citep{Penny2012} which might support including orphans, but given the limited extent of the disruption likely only $\sim90\%$ of the dark matter halo has been stripped \citep[e.g.,][]{Penarrubia2008}, and so we would not expect such a subhalo to be fully disrupted in TNG100. For each infall event, we record the satellite's M\textsubscript{200c} value from the FoF catalog of the snapshot prior to infall, when the satellite was its own central, and refer to this value as the infall mass. We note that using the \textsc{subfind} mass and converting it to M\textsubscript{200c} as we did for the hosts works equally well here, but we prefer to take the M\textsubscript{200c} directly from the FoF catalogs for the satellites as it avoids assuming a halo concentration. This choice of infall mass does not meaningfully affect our conclusions. Once these infall masses have been recorded, analogs of Antlia and Antlia B are chosen based on the 99.7\% credible intervals for each satellite, given in Table \ref{Table:m_halo_constraints}. Our final sample of Antlia analogs consists of over 5000 subhalos, while we identify over 20000 analog subhalos for Antlia B.\par

\begin{table}
  \caption{The expectation values of M\textsubscript{200c} and infall mass selection ranges for analogs from IllustrisTNG derived from Equation \ref{equation:halomassexpectation}, compared to the expectations from a naive inversion of the SMHM relation neglecting scatter. The selection ranges enclose 99.7\% of the halo mass PDFs.}
  \centering
    \hspace*{-4em}
    \begin{tabular}{c c c c c}
      \hline
      \hline
       Object & $\langle$M\textsubscript{200c}$\rangle$ & Naive Inversion & Lower Limit & Upper Limit\\
        & $[10^9 \, \text{M}_\odot]$ & $[10^9 \, \text{M}_\odot]$ & $[10^9 \,  \text{M}_\odot]$ & $[10^9 \, \text{M}_\odot]$ \\
      \hline
      NGC 3109 & 38.7 & 43.0 & 19.2 & 69.3 \\
      Antlia & 10.8 & 12.2 & 5.70 & 19.2 \\
      Antlia B & 3.12 & 4.26 & 1.26 & 8.13 \\
      \hline
    \end{tabular}
  \label{Table:m_halo_constraints}
\end{table}

\subsubsection{A Joint Sample} \label{subsubsection:jointsample}
For inclusion in our final sample, we only require that an NGC 3109 analog have \emph{either} an Antlia analog or an Antlia B analog. We formed a separate sample where we required each NGC 3109 analog to have \emph{both} an Antlia and an Antlia B analog, but found that the important quantities derived from the simulations (e.g., infall time distributions and orbital trajectories) were consistent between both samples. This ``joint" sample contained $\sim2500$ systems; given that there were 5000 systems with an Antlia analog and 20000 systems with an Antlia B analog when we required only one or the other, we may naively estimate that $\sim50\%$ of systems like NGC 3109 with an Antlia-like satellite also host a satellite like Antlia B, while $\sim12.5\%$ of systems with an Antlia B analog also host an Antlia analog. This indicates that, given the presence of Antlia, it is fairly common to find a satellite like Antlia B as well. We additionally find that, even though the infall time distributions of our Antlia and Antlia B analogs are similar (see \S\ref{subsubsection:infalltimes}), they rarely fell into their present-day hosts at the same time. Only $\sim5\%$ of systems in the ``joint" sample had their Antlia and Antlia B analogs fall into the NGC 3109 analog host in the same simulation snapshot, and in only $\sim0.5\%$ of the ``joint" samples were Antlia and Antlia B associated in the snapshot prior to infall. Given that the quantities relevant for our calculation are consistent between the separate and ``joint" samples, and that it is unlikely that Antlia and Antlia B fell in together, we choose to proceed by allowing NGC 3109 analogs to have either an Antlia or an Antlia B analog to increase our sample size, and we sort the satellites according to whether $P(\text{M}_h | \text{M}_*)$ (Equation \ref{equation:halomassprob}) is larger for the observed stellar mass of Antlia or Antlia B, given the satellite's halo mass at infall. In cases where an NGC 3109 analog includes multiple satellites with infall halo masses in our acceptance range (see Table \ref{Table:m_halo_constraints}), all such satellites are used in our analysis. \par

\subsubsection{Propagating Probabilities}
The formalism for calculating the probability that a simulated analog is representative of an observed satellite has the additional benefit of enabling propagation of these probabilities through further analysis on the simulated analogs. For a quantity $X$ measured from the simulated analogs (e.g., infall time), we define $x_i$ to be the value of $X$ measured for subhalo $i$. We can derive an improved estimate of the value of $X$ for the observed system by weighting each $x_i$ by the probability that the simulated system is representative of the observed system. Denoting these weights as $w_i$, we find

\begin{equation}
  \begin{aligned} \label{equation:simprob}
    w_i&=P(\text{M}_{h,\text{sat},i}|\text{M}_{*,\text{sat}}) \ P(\text{M}_{h,\text{host},i}|\text{M}_{*,\text{host}}) 
  \end{aligned}
\end{equation}

\noindent where $P(\text{M}_h|\text{M}_*)$ is defined as in Equation \ref{equation:halomassexpectation}. Estimates can then be made for, e.g., the mean value of $X$ for the observed system as

\begin{equation}
  \overline{X}= \frac{ \sum_i w_i \, x_i}{\sum_i w_i} \\
\end{equation}

We can incorporate additional data, in particular the projected separation of the satellite and the host and the line-of-sight velocity difference between the two, to further constrain the simulated analog set. This has historically been done by Monte Carlo rejection sampling, but it can be done more efficiently; in Appendix \ref{appendix:pq} we derive analytic forms for the PDFs of these projected quantities for simulated systems with full 6D information. These probabilities are multiplied into the weights defined in Equation \ref{equation:simprob} to give the final weights for each simulated analog (Equation \ref{equation:simprob2}). 

\section{Gas Mass Evolution} \label{section:massevolution}
In this section, we describe the semi-analytic framework we develop to assess the relative importance of different quenching mechanisms in terminating star formation in satellite galaxies. The primary mechanisms we assess are starvation (e.g., the cessation of cold gas inflows after infall; \citealt{Larson1980,Peng2015}) and ram pressure stripping (RPS; \citealt{Gunn1972}). These are the main environmental processes thought to quench star formation of dwarfs after infall (see \citealt{Cortese2021} for a review). \par

In our semi-analytic model, we evolve the \HI gas masses of satellites along their orbits in the host potential via a set of coupled ODEs. Once a satellite enters the host halo, with galaxy properties set at infall, we treat the satellite as being starved of gas inflows. In the starvation model, gas is continuously removed from the satellite according to stellar feedback. We treat this gas removal as instantaneously proportional to the star formation rates (SFRs), as inferred from the measured SFHs of the satellites assuming an effective mass-loading factor $\eta$ for the stellar outflows ($d\text{M}_\text{gas}/dt \propto -\eta \, d\text{M}_*/dt$). Unlike other semi-analytic models, we treat RPS as a continuous process, and not as an instantaneous process at pericenter.  Because of this, and because of the relatively high mass ratio between the satellite and host halos, we model the satellite orbits explicitly, including dynamical friction. In this section, we describe how we model the initial conditions and quenching mechanisms and motivate our fiducial parameter choices for our semi-analytic model. \par

We start in \S\ref{subsection:ic} with a description of the initial conditions for the host and satellite analogs prior to infall. In addition to the satellite stellar masses (which we infer based on the measured SFHs of Antlia and Antlia B) and their infall dark matter halo masses (determined as in \S\ref{subsubsection:so}), we must specify the dark matter halo concentrations for the orbital integrations, as well as the \HI gas surface density profile of the satellite, $\Sigma_\HI(r)$, and the density profile of the circumgalactic medium (CGM) of the host, $\rho_{\text{host}}(R)$, for the RPS calculation. Care in modeling the gas distributions is particularly important for testing RPS as a quenching mechanism.  \par

In \S\ref{subsection:quench}, we describe the physics and our specific semi-analytic implementations of starvation and RPS as post-infall quenching mechanisms of satellite star formation. We ignore any quenching process that might begin prior to infall onto the NGC 3109 analog host. This lack of pre-processing is justified by the simulations of \citealt{Jahn2021}, who find it unimportant for satellites with stellar masses similar to the Antlias and hosts similar in mass to NGC 3109. We describe our model for each mechanism and motivate our specific parameter choices, and show how we evolve the satellite orbits through the potential of the host with an analytic model for dynamical friction. \par

Throughout this section we will focus on our fiducial model while pointing out parts of the model that are uncertain. Results for our fiducial model are presented in \S\ref{section:results}. An exploration of alternate model choices is presented in \S\ref{subsection:am} and a deeper discussion of model uncertainties is presented in \S \ref{subsection:uncertainties}.\par

\subsection{Initial Conditions} \label{subsection:ic}
All integrations of our semi-analytic ODEs are initialized at the lookback time corresponding to the first snapshot in which a satellite was recognized as a subhalo of the host. Important initial conditions include the halo masses of the host and satellite, the stellar mass of the satellite at infall, the CGM density of the host, and the total mass and surface density profile of atomic hydrogen in the satellite.\par

We take dark matter halo properties (e.g., mass and infall velocity vector) directly from the analog sample described in \S\ref{subsection:analogselection}.  The initial halo mass for the host is taken to be the converted M$_{\text{200c}}$ inferred from the \emph{SubhaloMass} column of the simulation catalogs, as discussed in \S\ref{subsubsection:so}. For the satellite, we set the initial halo mass to be the FoF M$_{\text{200c}}$ in the snapshot prior to infall to avoid assuming a halo concentration.  We assume the stellar mass of the satellite at infall is fixed according to the measured SFHs of the satellites from \cite{Hargis2020} for Antlia B and \cite{Weisz2011} for Antlia. \par

We model the density of the host CGM as a singular isothermal sphere ($\rho \propto r^{-2}$), as suggested by simulations \citep[e.g.,][]{Fielding2017,Hafen2019a}, with a density normalization of $n=10^{-3}$ cm$^{-3}$ at 0.1 R$_{\text{vir}}$, following the fiducial high $\eta$ results from \cite{Fielding2017} (see their table 1). We note that the lightest halo considered by \cite{Fielding2017} has a mass of $10^{11}$ M$_\odot$, with an overdensity criterion of 200 times the mean density of the Universe (i.e., M$_{200\text{m}}$), while our estimated halo mass for NGC 3109 is only $3.87\times10^{10}$ M$_\odot$ based on a density criterion of 200 times the critical density of the Universe (i.e., M$_{200\text{c}}$). From their figure 7, it is clear the density profile of the CGM, even as a function of $r/\text{R}_{\text{vir}}$, evolves with halo mass, especially as the halos become less massive. Thus this normalization has an uncertainty which we discuss in more detail in \S\ref{subsubsection:hostcgm}.\par

As we integrate the satellite through the host's potential (described in \S\ref{subsection:quench}), we do not evolve the masses of the dark matter halos through the ODE integration, so neither can the host CGM density be evolved. We therefore choose to use the present-day value of the host virial radius to set the density normalization, such that RPS at earlier times will be more effective than in a self-consistently evolved calculation. This has a minimal effect for recent infalls, and we will show that other components of our model (principally, the satellite \HI surface density profile) limit the effectiveness of RPS for early infalls, even with this simplification. For our fiducial host halo mass, this corresponds to a cumulative CGM mass within 100 kpc of $\sim10^9$ M$_{\odot}$. While generally high $\eta$ models produce galaxies with clumpier circumgalactic media compared to low $\eta$ models, we do not attempt to model CGM clumpiness, which would effectively add stochasticity to our gas mass evolutions \citep{Simons2020,Akins2021}.\par

For the initial gas masses of the satellites, we use the double-power-law fit of M\textsubscript{\HI} to M$_*$ from \cite{Bradford2015}, based on measurements of isolated galaxies selected from the NASA Sloan Atlas \citep{Blanton2011,Geha2012}. Similar fits for samples that are selected via radio or infrared luminosities \citep[e.g.,][]{Scoville2017} generally prefer higher gas masses at lower stellar masses, but are likely biased in this regime due to completeness effects. Typical infall gas masses are $\sim1.5\times10^6$ M$_{\odot}$ for Antlia B analogs and $\sim1.7\times10^7$ M$_\odot$ for Antlia analogs. We do not include redshift evolution of the \HI mass scaling, as theoretical work predicts very weak redshift scaling of the M$_*$ -- M$_\HI$ relation out to $2<z<3$ \citep[e.g.,][]{Popping2015} and there are no observational constraints for such low-mass galaxies at these redshifts. We calculate the stellar masses of the satellites at infall by integrating the SFHs; thus satellites with earlier infalls have lower initial stellar and gas masses. We assume constant SFRs between bins in the SFHs \citep{Weisz2011,Hargis2020}.\par

For the satellite \HI distributions, we assume an exponential surface density profile $\Sigma_\HI(r) = \Sigma_0 \, \text{exp}(-r/r_s)$ as is observed over a wide range of M$_\HI$. As galaxies are observed to follow a tight locus in D$_\HI$ (defined as the \HI diameter where the surface density equals 1 M$_{\odot}$ pc$^{-2}$) and M$_\HI$, we set the initial \HI scale radii following this relation, which is given as $\text{log}_{10} \, \text{D}_\HI=0.506 \, \text{log}_{10} \text{M}_\HI - 3.293$ by \cite{Wang2016}, neglecting uncertainties.  Typical values of $r_s$ are $\sim250$ pc for Antlia B analogs and $\sim750$ pc for Antlia analogs. The gas scale radius is similar to the stellar half-light radius for Antlia B analogs, while the gas scale radius is about 50\% larger than the stellar half-light radius for Antlia analogs.  We find typical $\Sigma_0$ values of $4.0$ M$_{\odot}$ pc$^{-2}$ for Antlia B analogs and $4.3$ M$_{\odot}$ pc$^{-2}$ for Antlia analogs. \par

\subsection{Quenching Models}\label{subsection:quench}
In this section we show how we semi-analytically model quenching processes for our analog NGC 3109 systems.  In summary, our quenching model is formulated as a system of ODEs with the following components:
\begin{enumerate}
    \item Two-body orbit integration of the satellite and host after first infall, including dynamical friction (\S\ref{subsubsection:dynamicalfriction}). 
    \item RPS due to the gaseous halo of the host (\S\ref{subsubsection:RPS}), which depends on the orbit models in \S\ref{subsubsection:dynamicalfriction}.
    \item Starvation in the satellite due to cessation of gas inflows upon infall and mass loss due to star-formation-driven outflows (\S\ref{subsubsection:starvation}).
\end{enumerate}

\subsubsection{Satellite orbits and dynamical friction} \label{subsubsection:dynamicalfriction}
The modeling of RPS in particular requires that we track satellite orbits through the host potential.  The time resolution of the IllustrisTNG snapshots is not sufficient to resolve orbits, so we develop the following model to trace orbits to our desired resolution.  As noted in \S\ref{subsubsection:hmp}, our analysis indicates that Antlia was likely $\sim25\%$ as massive as NGC 3109 at infall. Thus it is expected that Antlia's infall will induce some reflex motion of the center of NGC 3109's halo -- this invalidates the assumption of a static host potential required for an analytic pericenter estimation, as was employed in \cite{Garling2020}. Therefore, we must track the orbital evolution post-infall to determine a reliable pericenter for considering RPS. We take this one step further: by formulating the orbital evolution of the host and satellite as a system of ordinary differential equations (ODE), we can additionally couple the quenching mechanisms directly to the orbit.  \par 

We implement the host and satellite system as a system of two rigid, extended bodies (as \citealt{Gomez2015} did to study the interaction of the LMC with the Milky Way). We use NFW \citep{Navarro1996} density profiles for the dark matter halos of the galaxies and use the median concentration-mass relation of \cite{Diemer2019a} as calculated for our adopted Planck 2015 \citep{Planck2016} cosmology to set the scale radii. We set the halo masses of both objects to be constant over the ODE integration and equal to their halo masses at the satellite's first infall. We neglect gravitational forces from the baryonic components of the galaxies, as the stellar-to-halo mass ratios of the satellites are low ($\sim 2\times10^{-4}$ for Antlia B and $\sim 7\times10^{-4}$ for Antlia). \par

The system of rigid bodies we have constructed does not experience dynamical friction, which can decrease the pericenter distances of the satellites and thus increase the effectiveness of RPS. It is important that we include dynamical friction because the satellite dark matter halos are comparable in mass to those of their hosts (see Table \ref{Table:m_halo_constraints}). We add this effect to the satellite only, using the standard approximation \citep{Chandrasekhar1943,Binney2008}

\begin{equation}
  \begin{aligned}
    \frac{d\mathbf{V}_{\text{sat}}}{dt} &= -4 \pi G^2\text{M}_{\text{sat}}\rho_{\text{host}}(R) \, \text{ln}\Lambda \times \\
    & \left[\int_0^{|\mathbf{V_{\text{sat}}}|} v^2 f_{\text{host}}(v)dv \right] \frac{\mathbf{V}_{\text{sat}}}{|\mathbf{V}_{\text{sat}}|}^3
  \end{aligned}
\end{equation}

\noindent where $f$ is the velocity distribution function and $\Lambda$ is the Coulomb factor. Assuming a Maxwellian velocity distribution, the integral can be approximated as

\begin{equation}
  \int_0^{|\mathbf{V_{\text{sat}}}|} v^2 f_{\text{host}}(v)dv \approx \erf{x} - \frac{2x}{\sqrt{\pi}} \, \text{exp}\left(-x^2\right)
\end{equation}

\noindent where $x=|\mathbf{V}_{\text{sat}}|/\sqrt{2 \, \sigma^2}$ and $\sigma$ is the one-dimensional velocity dispersion of the host's dark matter halo. The velocity dispersion can be approximated \citep[e.g.,][]{Zentner2003} or calculated by solving the Jeans equation

\begin{equation}
\frac{1}{\rho} \frac{d \, (\rho \sigma_r^2)}{dr} + 2\beta \frac{\sigma_r^2}{r} = -\frac{d\Phi}{dr}
\end{equation}

\noindent where $\sigma_r$ is the radial velocity dispersion and $\beta(r)=1-\sigma_{\theta}^2(r) / \sigma_r^2(r)$ is a measure of the anisotropy in the velocity distribution. We adopt the solution for constant $\beta$ and set $\beta=0.4$ \citep{Lokas2001}. For the Coulomb factor, we use the semi-analytic model of \cite{Petts2015} where

\begin{equation}
  \begin{aligned}
  b_{\text{max}} &= \text{min} \left(\rho_{\text{host}}(R) \ / \ \frac{d \rho_{\text{host}}(R)}{dr}, R \right) \\
  b_{\text{min}} &= \text{max} \left(r_{\text{hm}}, G \, \text{M}_{\text{sat}} \ / \ |\mathbf{V}|^2 \right) \\
  \Lambda &= 
  \begin{cases}
    & \dfrac{b_{\text{max}}}{b_{\text{min}}} \ \text{if} \ b_{\text{max}}>b_{\text{min}} \\
    & 0, \ \text{otherwise} \\
  \end{cases}
  \end{aligned}
\end{equation}

\noindent where $b_{\text{max}}$ and $b_{\text{min}}$ are the maximum and minimum impact parameters, and $r_{\text{hm}}$ is the half-mass radius of the satellite.

\subsubsection{Ram Pressure Stripping} \label{subsubsection:RPS}
Although RPS was originally formulated in the context of hot gas halos \citep[e.g.,][]{Gunn1972,Tonnesen2009,Fillingham2016,Roberts2019}, evidence is mounting that the cool CGM can contribute significantly to RPS in low-mass hosts \citep{Roediger2005,Simons2020}. RPS proceeds when the ram pressure from the host's gas halo 
\begin{eqnarray} \label{equation:rpsbasic}
P_{\text{RPS}}(R)&=\rho_{\text{CGM}} (R) \ |\mathbf{V_{\text{sat}}}|^2 
\end{eqnarray}

\noindent exceeds the maximum gravitational restoring force per unit area of the satellite
\begin{equation} \label{equation:longrps}
\begin{aligned}
P_{\text{restore}}(r) &= \Sigma_{\text{gas,sat}} (r) \ \frac{G \, \text{M}_{\text{sat}} (r)}{r^2} \\
    &= \Sigma_{\text{gas,sat}} (r) \ \frac{d\Phi_{\text{sat}}(r)}{dr}. \\
  \end{aligned}
\end{equation}

\noindent for a system with dynamics dominated by a spherical potential, and can be written as
\begin{equation}
    P_{\text{RPS}}(R) > P_{\text{restore}}(r).
\end{equation}

\noindent Here, $R$ is the distance from the satellite to the host, $r$ is the distance from the point being considered in the satellite's disk to the satellite's center, $G$ is the gravitational constant, $\Sigma_{\text{gas,sat}}(r)$ is the gas surface density of the satellite at $r$, $\text{M}_{\text{sat}}(r)$ is the total mass of the satellite enclosed within radius $r$, $\Phi$ is the gravitational potential of the satellite, $\mathbf{V}_{\text{sat}}$ is the velocity vector of the satellite with respect to the host's gaseous halo (called simply \textbf{V} in the prior section), and $\rho_{\text{CGM}}(R)$ is the density of the host's gas halo at $R$ (\citealt{McCarthy2008}; see also \citealt{Koeppen2018}, which gives an alternate formulation for cases where a stellar disk potential is important). The value of $r$ at which $P_{\text{RPS}} = P_{\text{restore}}$ is called the stripping radius ($r_{\text{strip}}$), and is minimized at pericenter where the host gas halo density and satellite velocity are maximized. For this reason, RPS is often implemented as happening instantly at pericenter or occuring gradually over the relevant pericenter timescale \citep[e.g.,][]{Font2008}. However, not all satellites may experience a pericenter passage by present day. Moreover, most gradual RPS schemes neglect dynamical friction, which is likely to be important for the Antlia analogs, so we adopt a different method for calculating RPS.\par

To include RPS in our ODE, we require time differentials related to the satellite's orbital parameters in order to evolve the stripping radius and remaining H\textsc{i} mass dynamically. We write the time differential of the RPS pressure as

\begin{equation}
  \begin{aligned}
    \frac{d\,P_{\text{RPS}}}{dt} &= \frac{d\,\rho_{\text{CGM}}(R)}{dR} \, \frac{d\,R}{dt} \, |\mathbf{V}_{\text{sat}}|^2 \\
    & + 2 \, \rho_{\text{CGM}}(R) \, |\mathbf{V}_{\text{sat}}| \, \frac{d\,|\mathbf{V}_{\text{sat}}|}{dt}
  \end{aligned}
\end{equation}

\noindent where $dR/dt$ in the first term can be written as the scalar product $\mathbf{V}_{\text{sat}} \cdot \hat{\mathbf{R}}$ with $\hat{\mathbf{R}}$ being the unit vector in the direction of the host center. If we assume that the only significant bulk motion of the host's gaseous halo is the reflex motion due to the satellite we can rewrite $d|\mathbf{V}_{\text{sat}}|/dt$ as $\mathbf{a_{\text{sat}}}\cdot\hat{\mathbf{V}}_{\text{sat}}$, which gives the component of the acceleration in the direction of the velocity. Substituting these terms, we have

\begin{equation}
  \begin{aligned}
    \frac{d\,P_{\text{RPS}}}{dt} &= \frac{d\,\rho_{\text{CGM}}(R)}{dR} \, \left(\mathbf{V}_{\text{sat}} \cdot \mathbf{\hat{R}}\right) \, |\mathbf{V}_{\text{sat}}|^2 \\
    & + 2 \, \rho_{\text{CGM}}(R) \, |\mathbf{V}_{\text{sat}}| \ \left( \mathbf{a_{\text{sat}}}\cdot\hat{\mathbf{V}}_{\text{sat}} \right)
  \end{aligned}
\end{equation}

\noindent which contains only basic terms related to the orbit and the host gas density profile. With this expression, we can track $P_{\text{RPS}}$ explicitly in our ODE. In general this could be done in a post-processing step after orbit integration, but including the differential for $P_{\text{RPS}}$ is useful when using error-controlled, adaptive-timestep ODE integrators. 

Additionally, with a differential form for the ram pressure, we can also look for a differential form for the stripping radius. Generally the stripping radius must be solved numerically (for example, by root-finding) as the solution of $P_{\text{RPS}} = P_{\text{restore}}$ is rarely analytic. However, the equality $P_{\text{RPS}}(t) = P_{\text{restore}}(r_{\text{strip}})$ which holds at the stripping radius requires that 
\begin{equation}
dP_{\text{RPS}}/dt = dP_{\text{restore}}/dr_{\text{strip}} \times dr_{\text{strip}}/dt
\end{equation}
so that 
\begin{equation}
dr_{\text{strip}}/dt = \frac{dP_{\text{RPS}}/dt}{ dP_{\text{restore}}/dr_{\text{strip}}}. 
\end{equation}
\noindent We have already formulated the time differential for $P_{\text{RPS}}$, so now we need the radial differential for $P_{\text{restore}}$, which can be written as

\begin{equation}
  \begin{aligned}
    \frac{dP_{\text{restore}}}{dr} (r) &= \frac{d \, \Sigma_{\text{gas,sat}}}{dr}(r) \, \frac{d\Phi_{\text{sat}}}{dr}(r) \\
    &+ \Sigma_{\text{gas,sat}}(r) \, \frac{d^2 \Phi_{\text{sat}}(r)}{dr^2} \\
    \\
  \end{aligned}
\end{equation}

\noindent The time differential of $r_{\text{strip}}$ is then simply

\begin{equation}
  \frac{dr_{\text{strip}}}{dt} = \frac{dP_{\text{RPS}}}{dt} \ / \ \frac{dP_{\text{restore}}}{dr} (r_{\text{strip}})
\end{equation}

\noindent Thus, we must only compute the stripping radius numerically once to set the initial value, and we can then track its evolution via the differentials, which have components that are analytic for most gas distributions and potentials. We can then write the gas mass differential as

\begin{equation}
    \frac{d \, \text{M}_{\text{gas,sat,tot}}}{dt} =
    \text{min}\left(0, \, \dfrac{\displaystyle d \, \text{M}_{\text{gas,sat,enc}}}{dr_{\text{strip}}} \, \dfrac{\displaystyle r_{\text{strip}}}{dt}\right)
\end{equation}

\noindent where $\text{M}_{\text{gas,sat,enc}}(r)$ is the total gas mass of the satellite enclosed within radius $r$. This differential is always equal to or less than 0 by construction. \par

\subsubsection{Starvation} \label{subsubsection:starvation}

To define our model for starvation, we begin with our assumptions of the state of the atomic gas in the satellites prior to infall.  In isolation, it is typical in semi-analytic models to make gas accretion proportional to the halo mass growth rate \citep[e.g.,][]{Benson2012,Kravtsov2021}, with the simplest constant of proportionality being the cosmic baryon fraction, $\Omega_b / \Omega_m$. In field dwarfs, this accretion rate must outpace star-formation-driven outflows at early times to produce the high M\textsubscript{\HI} / M$_*$ ratios that are observed in field dwarfs (generally 1--6 at $\text{M}_*=10^8 \ \text{M}_{\odot}$; \citealt{Papastergis2012,Popping2015,Koribalski2018}). Given these high gas fractions, it is understandable that the vast majority of field dwarfs are observed to be actively star-forming \citep[e.g.,][]{Geha2012,Dickey2021}. \par

In starvation, it is assumed that these inflows are shut off after accretion to the host; this is typically explained by the presence of a hot gas halo in the host that intercepts these inflows. Recent simulations suggest that roughly LMC-mass hosts can sustain such a halo \citep[e.g.,][]{Jahn2021}, but generally the physical mechanism of starvation is uncertain for such low host masses. With gas accretion shut off, the dwarf can continue to form stars from its gas reservoir. However, this reservoir depletes over time due to star-formation-driven outflows. The gas supply eventually exhausts and star formation is quenched. We thus assume that starvation depends only on time since infall, the star formation rate, and outflows.  \par

We utilize a simple model for starvation, in which there is no net gas accretion to a satellite after infall, and the change in the gas mass is completely specified by

\begin{equation} \label{equation: basic_starvation}
  \frac{d \text{M}_{\text{gas,sat}}}{dt} = \left( R - 1 - \eta(t) \right) \, \frac{d\text{M}_*}{dt} 
\end{equation}

\noindent where $R$ is known as the recycled fraction and quantifies how much of the gas that goes into forming stars is returned to the ISM, and $\eta$ is the dimensionless mass-loading factor that relates the gas outflow rate to the SFR $\left(\frac{d\text{M}_*}{dt}\right)$. We adopt $R=0.3$ \citep{Portinari2004a}, but it makes little difference given our large mass-loading factors. To calculate the starvation rate, we utilize the SFH from \cite{Hargis2020} for Antlia B and \cite{Weisz2011} for Antlia. \par

The key parameter that most influences the importance of starvation is the mass-loading factor, and specifically how it varies as a function of galaxy or halo mass and/or time.  For our fiducial model, we adopt the power-law fit of \cite{Christensen2016}, which relates the mass-loading factor to the circular velocity at the virial radius as $\eta \propto v_{\text{circ}}^{-2.2}$. This fit is based on hydrodynamical simulations of galaxies with halo virial masses from $3\times10^9$ to $7\times10^{11}$ M$_\odot$ using the \textsc{gasoline} code \citep{Wadsley2004}. This halo mass range includes our expectations for both Antlia and Antlia B (see Table \ref{Table:m_halo_constraints}). Typical mass-loading factors for Antlia and Antlia B with this model are about 6 and 10, respectively. We explore alternate model choices in \S\ref{subsubsection:massloading}. \par

\begin{figure*}
  \centering
  \includegraphics[width=0.45\textwidth,page=1]{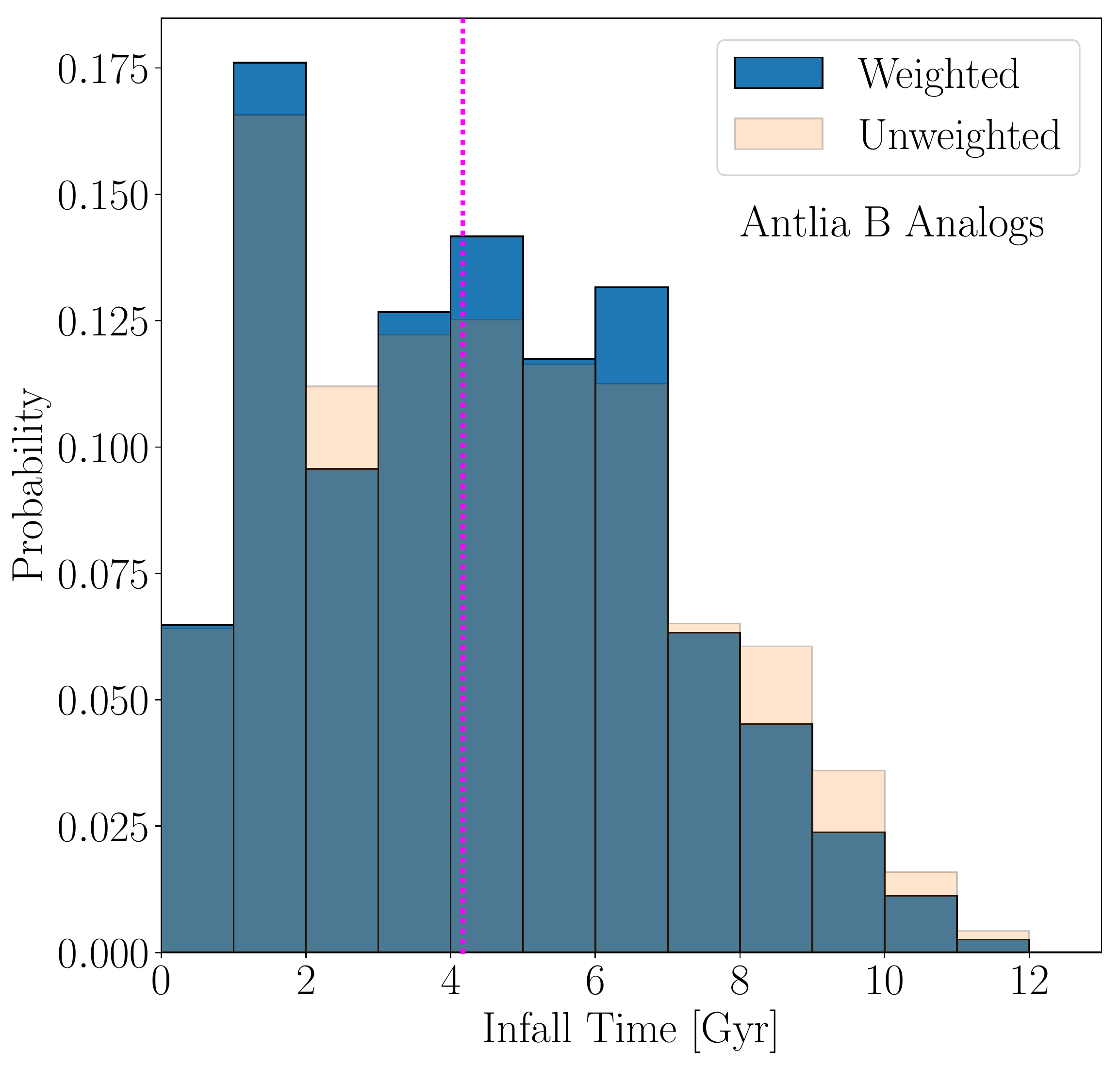}
    \includegraphics[width=0.45\textwidth,page=1]{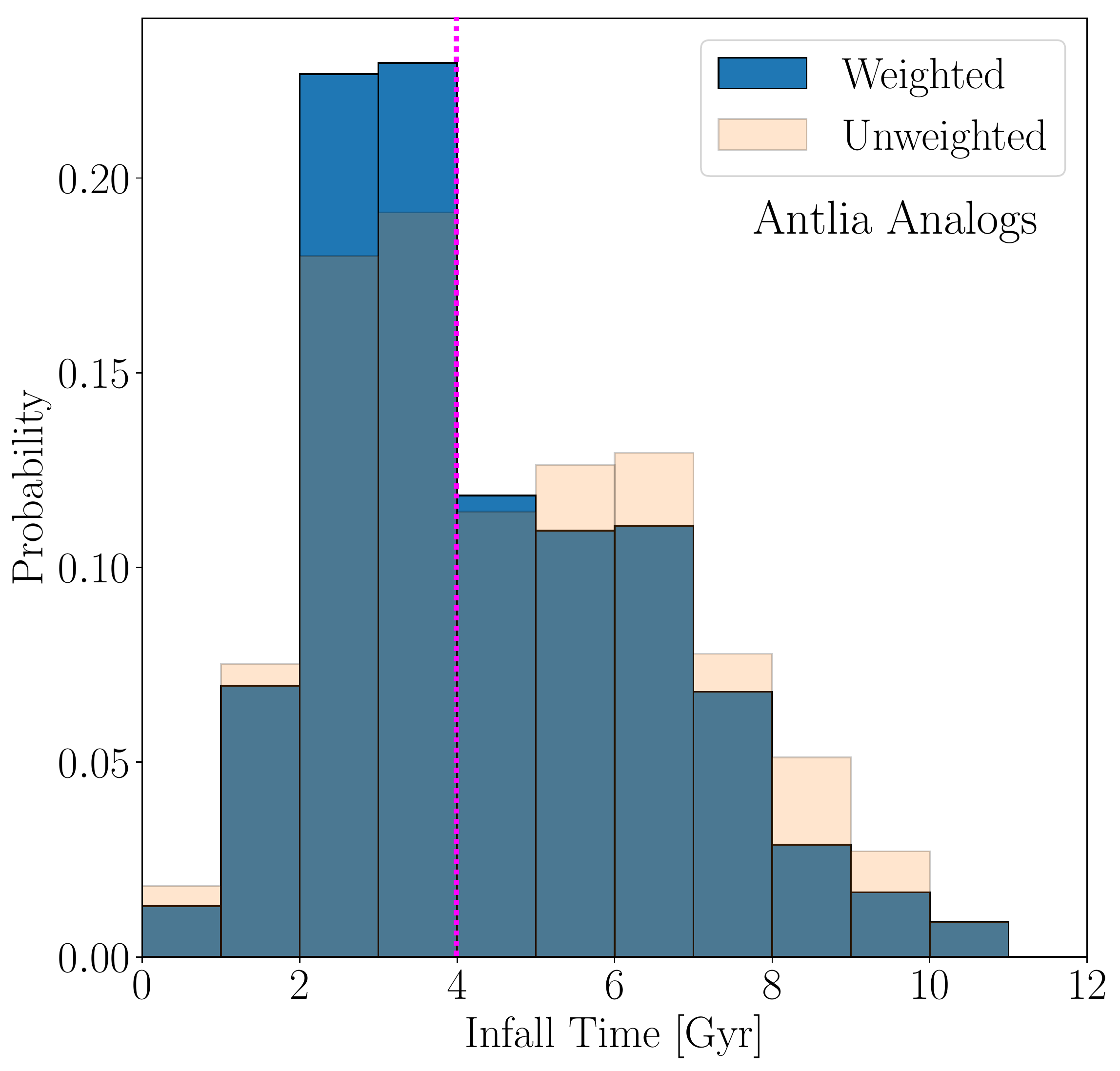}
  \caption{The probability distributions of lookback time at first infall (in Gyr ago) for Antlia B (left) and Antlia (right) analogs selected from TNG100. Weighted distributions are weighted by the host and satellite halo mass probabilities (\S\ref{subsubsection:hmp}) in addition to projected quantity probabilities (Appendix \ref{appendix:pq}), while unweighted distributions are weighted only by the projected quantity probabilities, as is more typical in the literature. The magenta dashed lines show the weighted median infall times for the analogs, which is about 4 Gyr ago for both satellites.}
  \label{figure:antlia_infalls}
\end{figure*}

\section{Results} \label{section:results}
In this work, we are primarily interested in assessing the relative efficacy of RPS and starvation in quenching satellites of low-mass hosts. This relative efficacy will change depending on the specific choices of the quenching parameters and initial conditions. For a given set of model parameters, the best way to assess the relative efficacy of RPS and starvation is to examine the gas mass evolutions of the set of simulated analog systems whose final, present-day \HI masses are consistent with those observed for the real satellite. \par 

Following this principle, we examine the relative efficacy of RPS and starvation under our fiducial quenching model in \S\ref{subsection:fiducialmodel}. We first present the distributions of infall times and pericenter distances for our full analog sample to provide insight into the typical properties of the simulated analog satellites. We then present the mean gas mass evolutions of our Antlia and Antlia B analogs in 1 Gyr bins of infall time under the effects RPS and starvation separately; this allows us to separate the effects of the two quenching mechanisms and illustrate the range of infall times which can reproduce the observed \HI masses of the satellites under each. \par 

We then proceed to examine the gas mass evolutions of individual samples under the effects of both RPS and starvation simultaneously to illustrate the variety of evolutionary paths that our satellite analogs take under our fiducial quenching model. In \S\ref{subsection:am} we examine how our results change under alternate quenching models, and in \S\ref{subsection:uncertainties} we show how observational and theoretical uncertainties affect our conclusions. \par

\begin{figure*}
  \centering
  \includegraphics[width=0.45\textwidth,page=1]{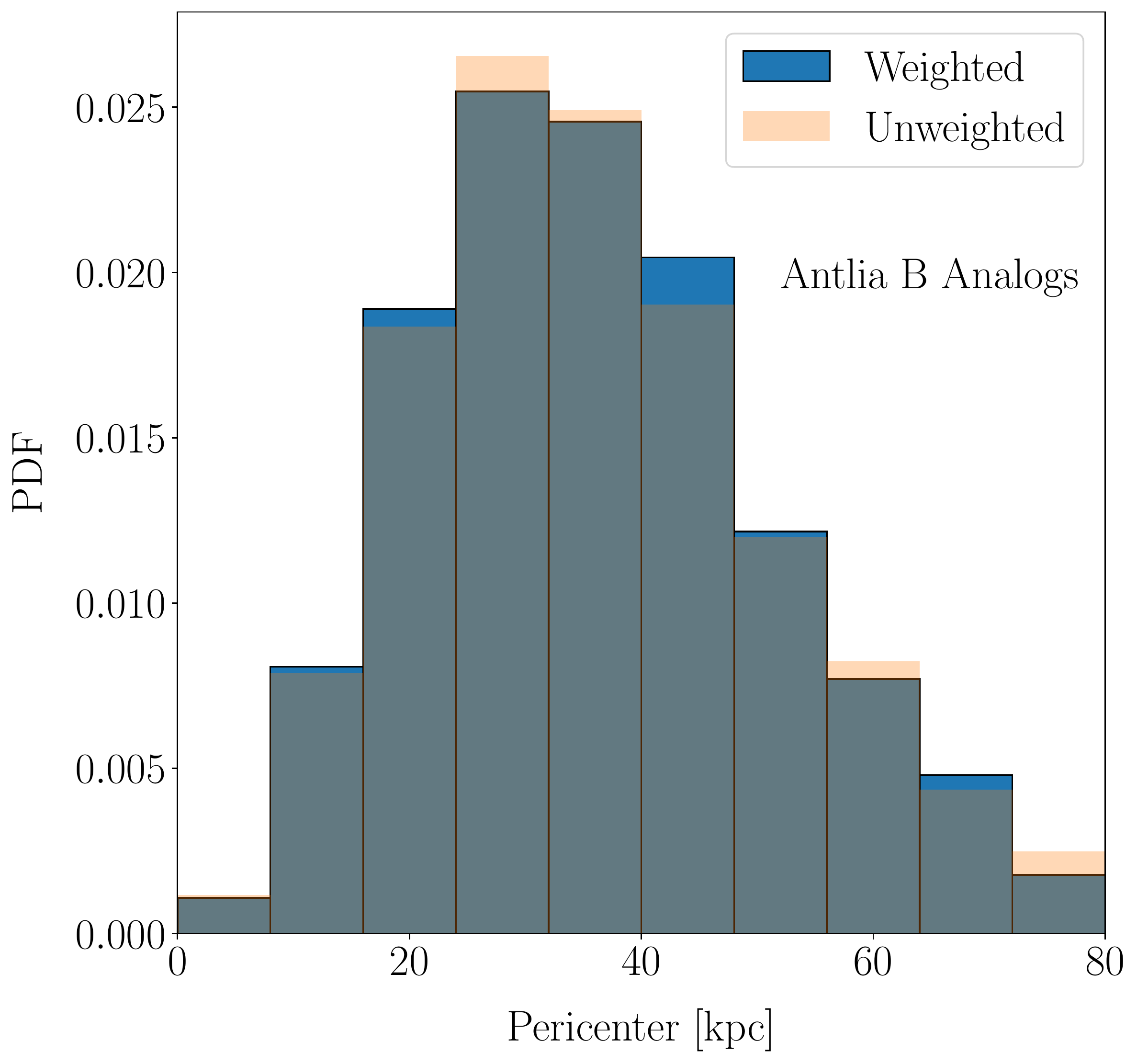}
  \includegraphics[width=0.45\textwidth,page=1]{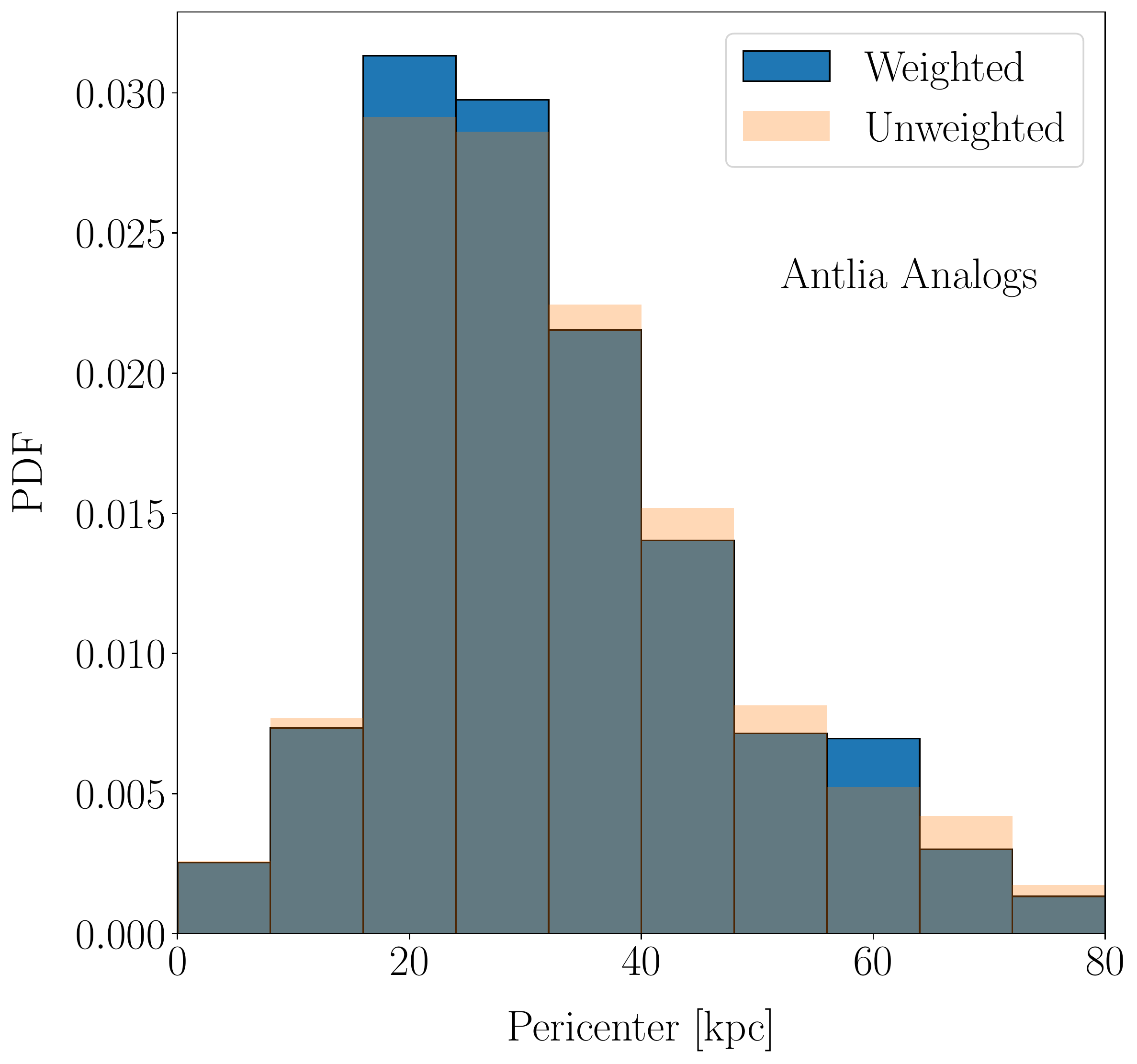} 
  \caption{Probability distributions for the host-satellite separation at first pericenter after infall for Antlia B (left) and Antlia (right) analogs selected from TNG100, with weights as in Figure \ref{figure:antlia_infalls}. The distributions are similar, with peaks around $\sim30$ kpc. About $7\%$ of the analog satellites experience pericenters $\leq10$ kpc where tidal stripping may become important (see \S\ref{subsubsection:tidalstripping}).}
  \label{figure:antlias_orbit}
\end{figure*}

\subsection{Fiducial Model} \label{subsection:fiducialmodel}
In this section we will study the evolution of Antlia B and Antlia analogs from the TNG100 simulations under the fiducial model choices presented in \S\ref{subsection:ic}. To facilitate interpretation, we focus on computing expected values and general trends and thus neglect uncertainties in the initial conditions and scaling relations used to compute the gas mass evolution. We discuss some of these sources of uncertainty in \S\ref{subsection:uncertainties} and find that the uncertainties in the absolute quenching timescales are quite large, typically spanning multiple Gyr. However, our goal is not to robustly estimate the infall times of Antlia and Antlia B using quenching timescales, but to assess the relative importance of starvation and RPS in depleting the gas reservoirs of these systems, and such relative comparisons are robust to these uncertainties.\par

Throughout the rest of the paper, weighted distributions utilize the full weights, including halo mass probabilities (\S\ref{subsubsection:hmp}) and projected quantity probabilities as defined in Equation \ref{equation:simprob2}, while unweighted distributions include only the probabilities of projected quantities as derived in Appendix \ref{appendix:pq}, to facilitate comparison to other work and illustrate the difference made by including halo mass probabilities. Generally, including halo mass probabilities makes a 10--20\% difference per bin across most distributions. We allow the satellite gas masses to go negative in our ODE integration to facilitate easier comparisons between different infall times, though we mark regions of negative gas masses as unphysical. We begin by discussing the infall time distributions, as these set the relevant quenching timescales for the satellites. \par

\begin{figure*} 
  \centering
  \includegraphics[width=0.45\textwidth,page=1]{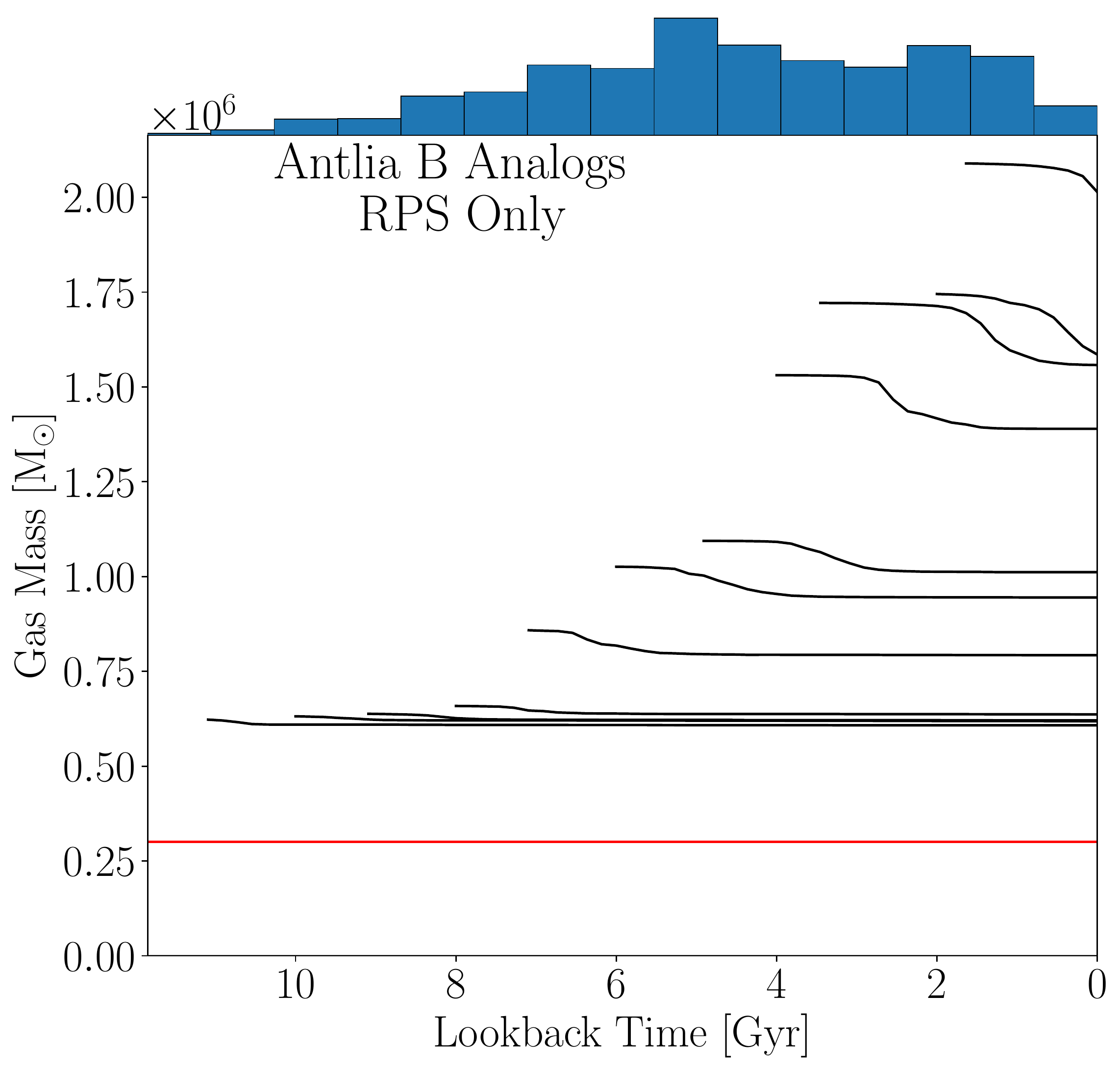} \hspace{1cm}
  \includegraphics[width=0.45\textwidth,page=1]{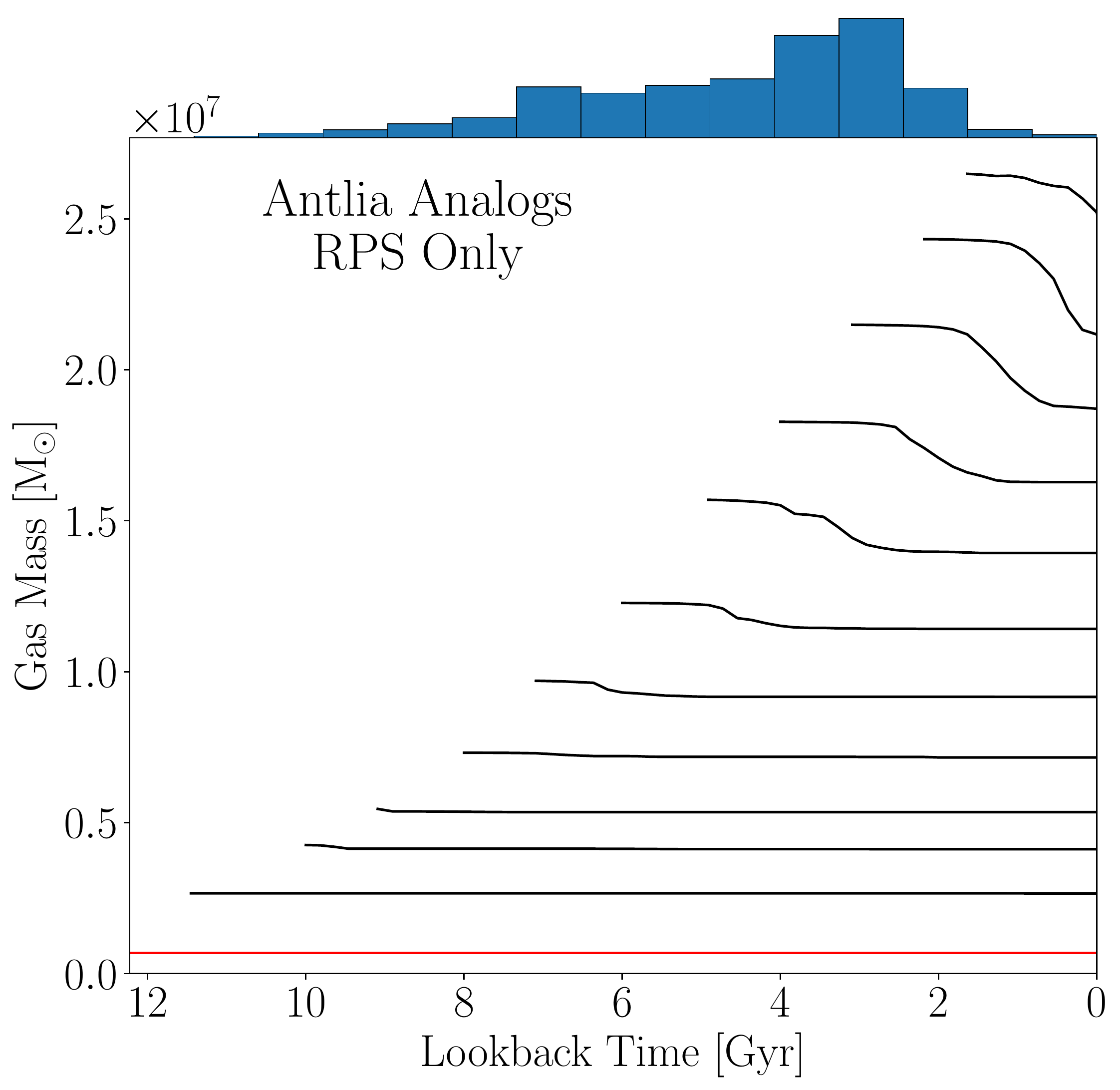} \\
  \includegraphics[width=0.45\textwidth,page=1]{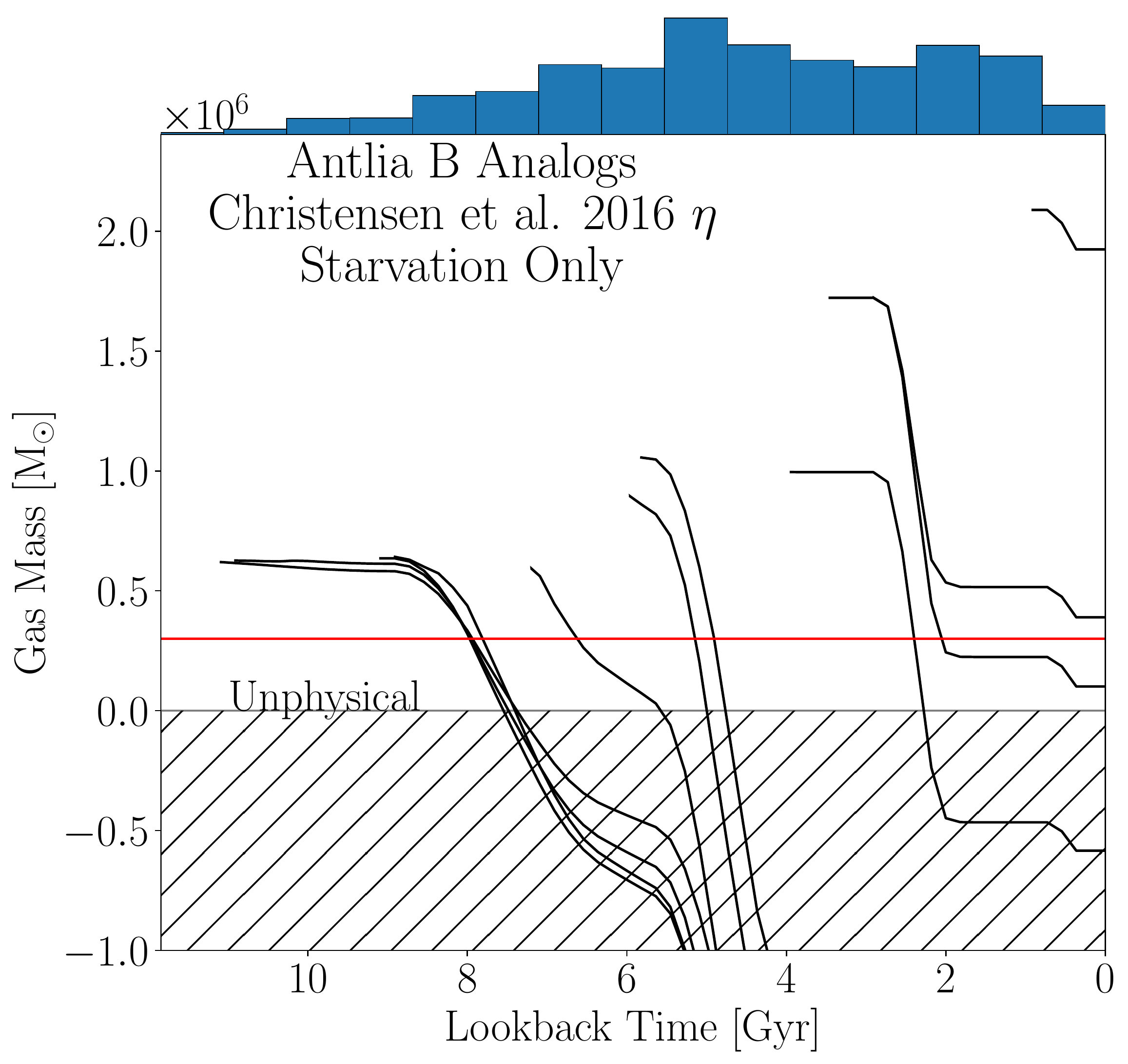} \hspace{1cm}
  \includegraphics[width=0.45\textwidth,page=1]{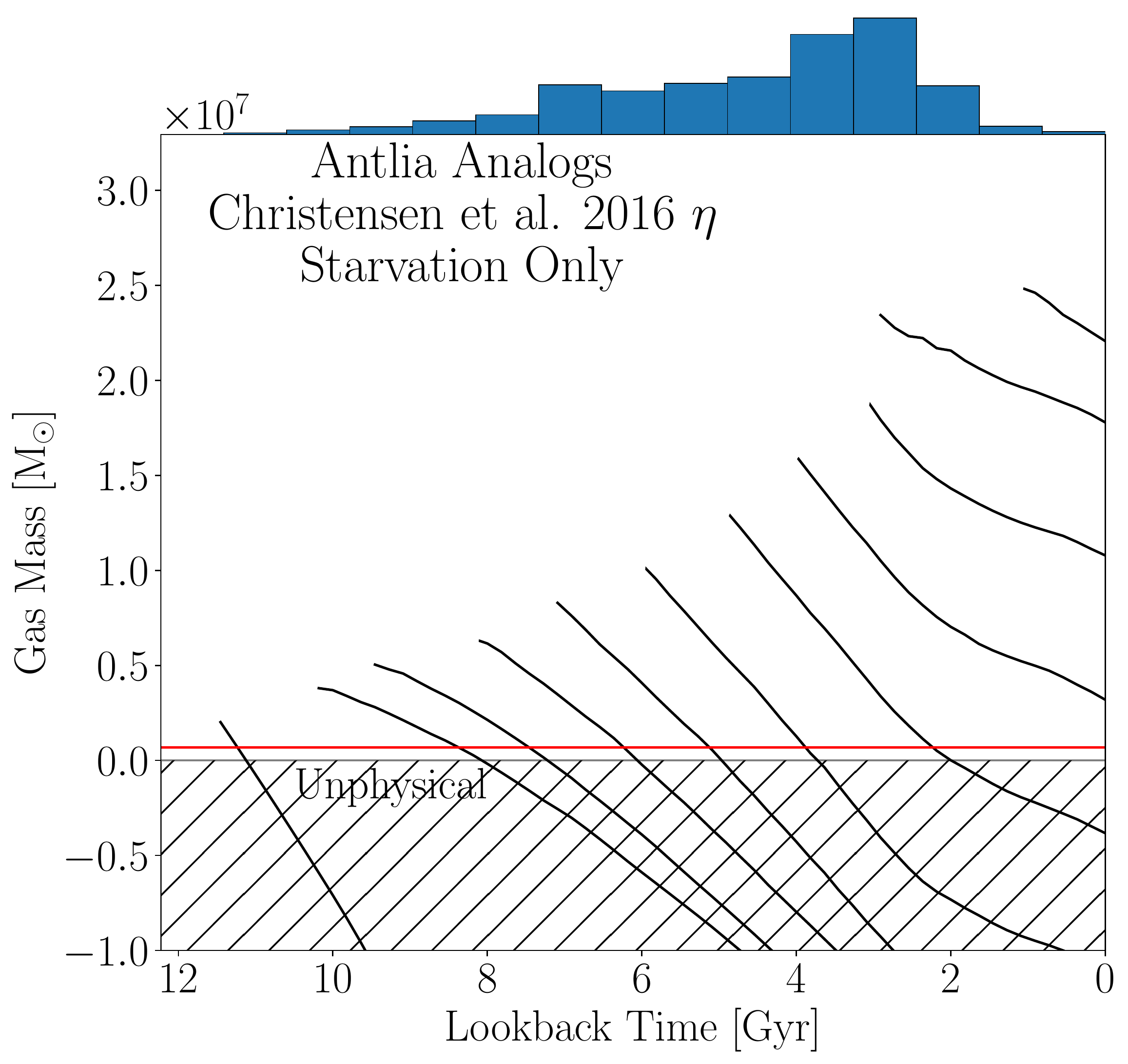}
  \caption{Gas mass evolutions for Antlia B analogs (left column) and Antlia analogs (right column) over the full range of analog infall times, considering RPS (top row) and starvation (bottom row) separately. The plotted tracks are weighted means across 1 Gyr bins of lookback time. Marginal histograms show the weighted infall time distributions from Figure \ref{figure:antlia_infalls}. Red horizontal lines mark the measured present-day gas masses of Antlia and Antlia B (see Table \ref{Table:systemproperties}). RPS is ineffective at early times due to our adopted \HI mass-size relation producing more compact gas distributions for lower infall gas masses. We find that RPS in our fiducial model produces present-day analogs which are too gas-rich across the entire range of infall times, while starvation can match the observed gas masses for infall times that are likely given the weighted infall time distributions.} 
  \label{figure:mean_evolution}
\end{figure*}

\subsubsection{Satellite Infall Times} \label{subsubsection:infalltimes}

We plot the infall time probability distributions for the Antlia B and Antlia analogs selected from TNG100 in Figure \ref{figure:antlia_infalls}. The distribution for Antlia B analogs shows a statistically significant peak in the infall time distribution between 1 and 2 Gyr ago, and relatively similar probability from 3--7 Gyr ago, with the probability dropping off for earlier infalls. Including the halo mass probabilities in the weights further disfavors earlier infall times. Antlia analogs show a slightly earlier peak in the infall distribution from 2--4 Gyr ago. Similar to Antlia B analogs, the Antlia analogs have fairly flat infall probability from 4--7 Gyr ago, with earlier infalls disfavored. Including the halo mass probabilities in the analysis for the Antlia analogs makes the peak at 2--4 Gyr ago more prominent and disfavors earlier infalls. Overall, the infall time distributions for the Antlia B and Antlia analogs show relatively similar patterns, but it is unlikely they fell into NGC 3109 together; see \S\ref{subsubsection:jointsample}. \par

\begin{figure*}
  \centering
  \includegraphics[width=0.45\textwidth,page=1]{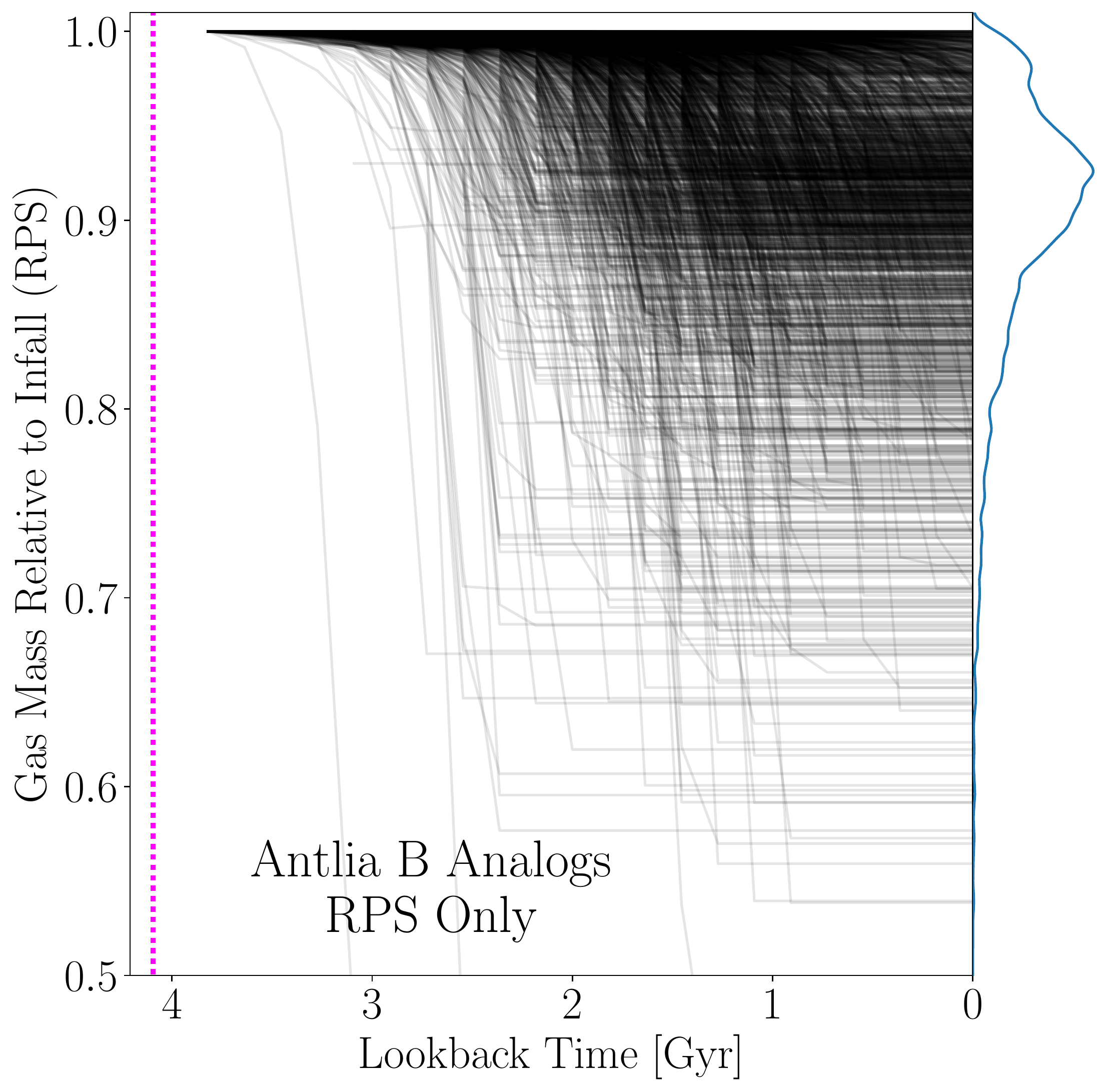}
  \includegraphics[width=0.45\textwidth,page=1]{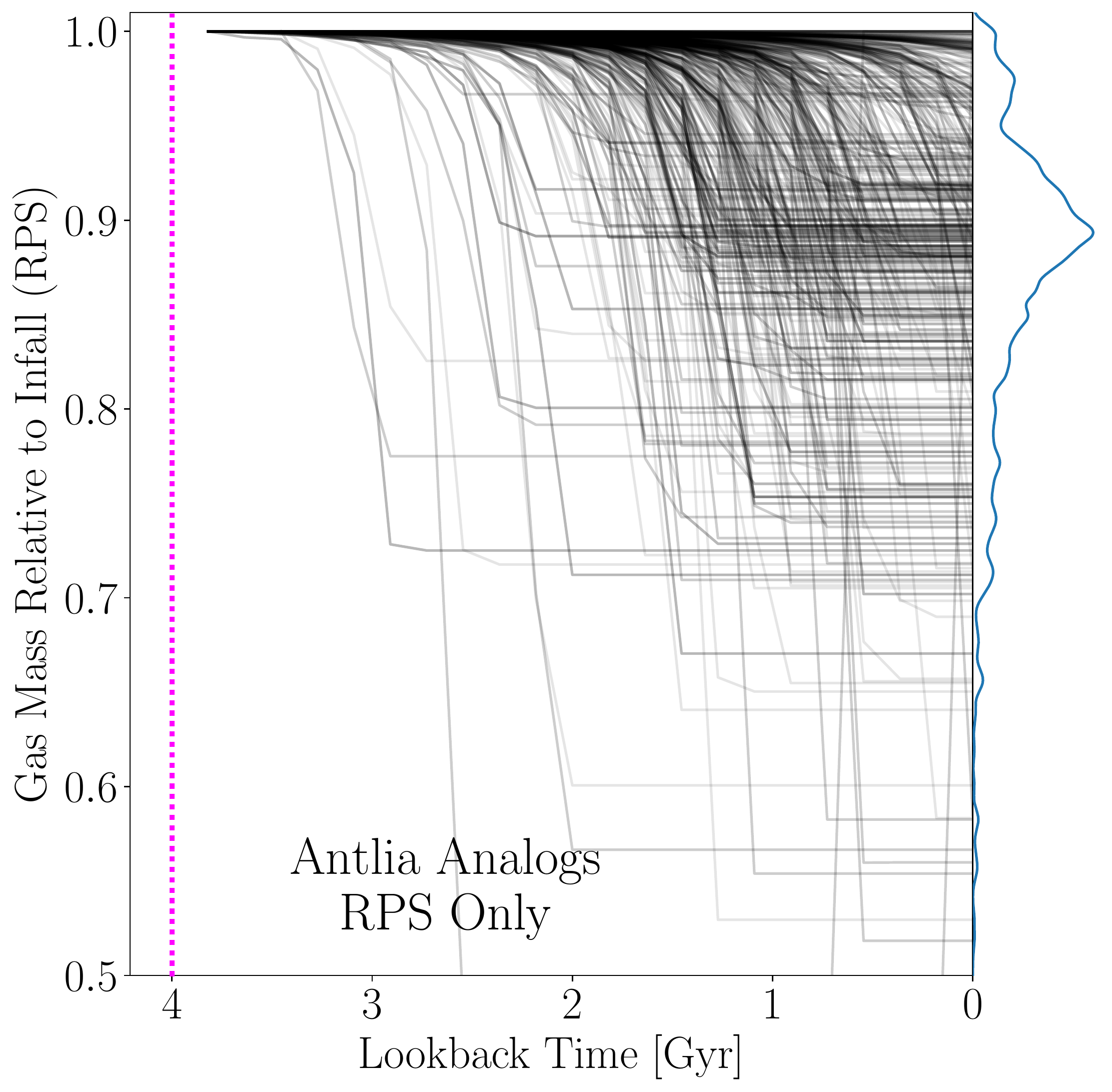} \\
  \includegraphics[width=0.45\textwidth,page=1]{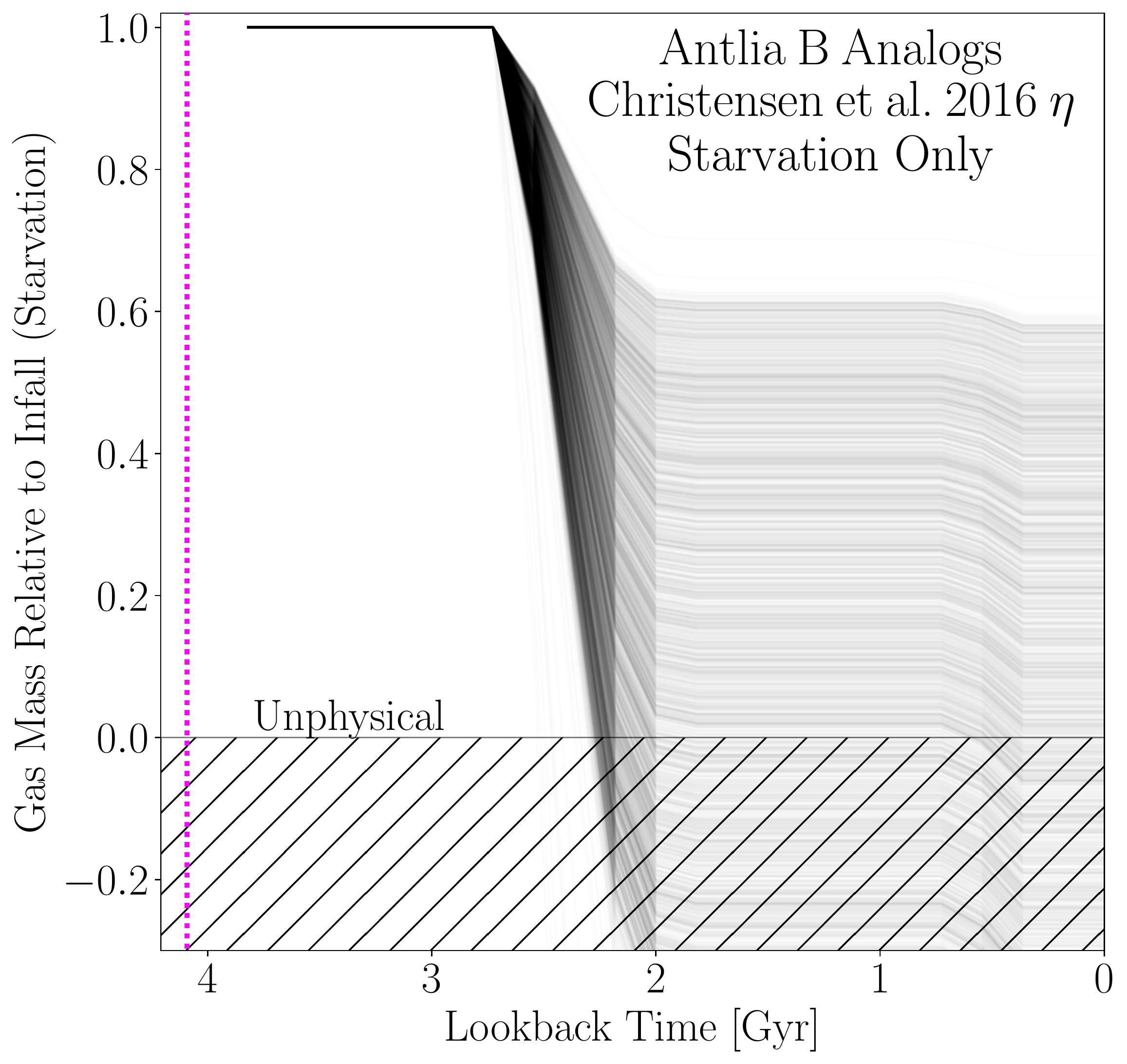}
  \includegraphics[width=0.45\textwidth,page=1]{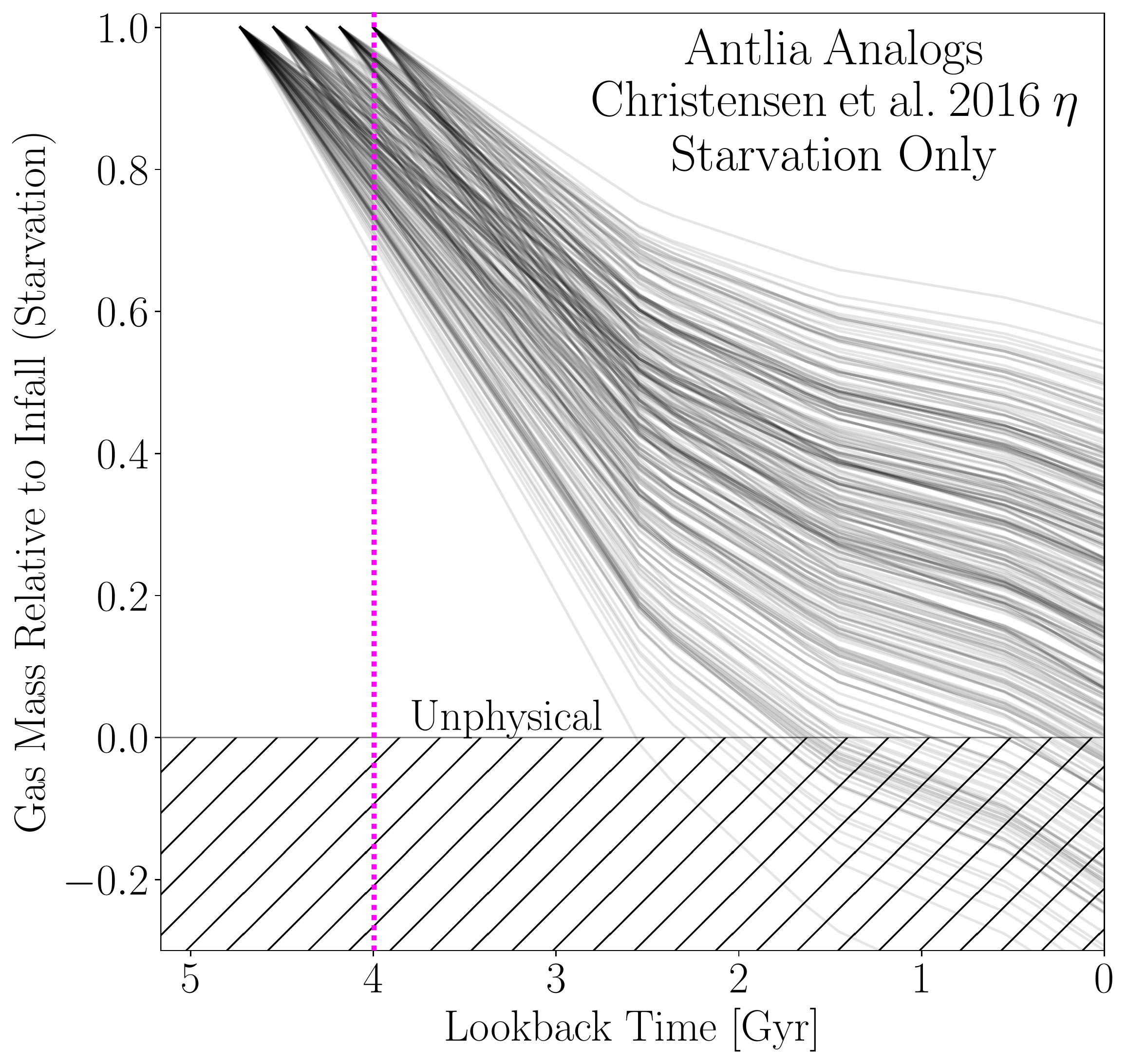}
  \caption{The relative gas mass evolution due to RPS (top row) and starvation (bottom row), calculated as the ratio of the gas mass as a function of time to the infall gas mass, for Antlia B (left column) and Antlia (right column) analogs in the ranges of infall times that give reasonable agreement with the observed \HI masses (generally 3--5 Gyr ago). The weighted median infall times for the full samples of Antlia and Antlia B analogs are indicated by vertical magenta dashed lines as in Figure \ref{figure:antlia_infalls}. The majority ($\sim95\%$) of analogs show less than $20\%$ gas mass loss due to RPS in this infall time range, while starvation removes the majority of the gas. The variation in the starvation plot is caused by differences in the satellite halo masses at infall.}
  \label{figure:relative_gas_evo}
\end{figure*}

\subsubsection{Orbital Parameters} \label{subsubsection:orbitalparameters}
As the pericenter distance of the satellites is important to the effectiveness of RPS, we show the distributions of first pericenters for Antlia B and Antlia analogs in Figure \ref{figure:antlias_orbit}. Only $\sim7\%$ of analog satellites experience close pericenters $\leq 10$ kpc for which tidal stripping may be important (see \S\ref{subsubsection:tidalstripping}). This may partly be due to survivor bias; we require satellites to survive until the present day in TNG100 to be selected, and satellites with small pericenters are more likely to be disrupted. The pericenter distributions reach their peaks at about 20--30 kpc. The means of the distributions are slightly larger than their modes as the distributions are mildly skewed to larger pericenters. Inclusion of halo mass probabilities affects the pericenter distributions minimally.\par 

\subsubsection{Mean Gas Evolution} \label{subsubsection:meanevolution}
In Figure \ref{figure:mean_evolution} we show the weighted mean gas mass evolutions in 1 Gyr bins of infall time for Antlia B (left column) and Antlia (right column) analogs, considering RPS (top row) and starvation (bottom row) separately. The marginal histograms show the infall time distributions from Figure \ref{figure:antlia_infalls} for each satellite, while the red horizontal lines mark the present-day observed \HI masses from Table \ref{Table:systemproperties}. \par

Under our fiducial model, RPS is never effective enough to reproduce the observed \HI masses of Antlia and Antlia B on its own. Even for early infall times, which should afford the satellites several pericenters over which to experience RPS, we find that RPS is made ineffective by our assumed \HI mass-size relation, which produces more compact gas distributions for lower infall gas masses; such compact distributions are quite resilient to stripping via RPS. Meanwhile, our fiducial starvation model is quite capable of reproducing the observed \HI masses for a feasible range of infall times. For Antlia B analogs, infalls in the range of 2--4 Gyr produce comparable \HI masses to those observed, and this infall time range is highly probable. For Antlia analogs, infalls in the range of 3--5 Gyr show good agreement with the observed \HI mass, which is again a preferred infall time range based on Figure \ref{figure:antlia_infalls}. \par

These results indicate that starvation better explains the quenching of the Antlias than RPS. It is also worth looking at how individual systems evolve, and how starvation and RPS act together; we explore this in the next section. \par

\begin{figure*}
  \centering
  \includegraphics[width=0.45\textwidth,page=1]{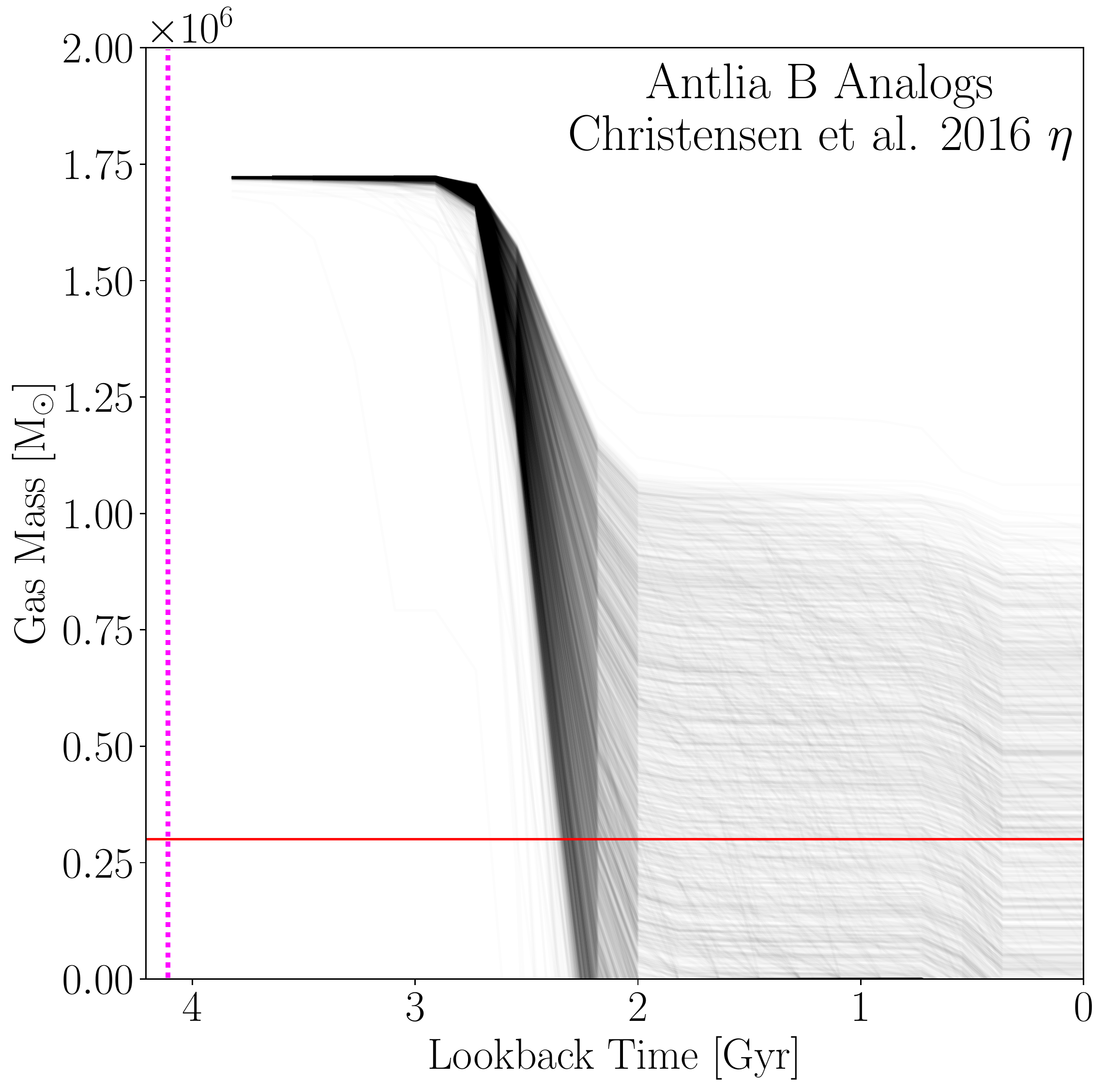}
    \includegraphics[width=0.45\textwidth,page=1]{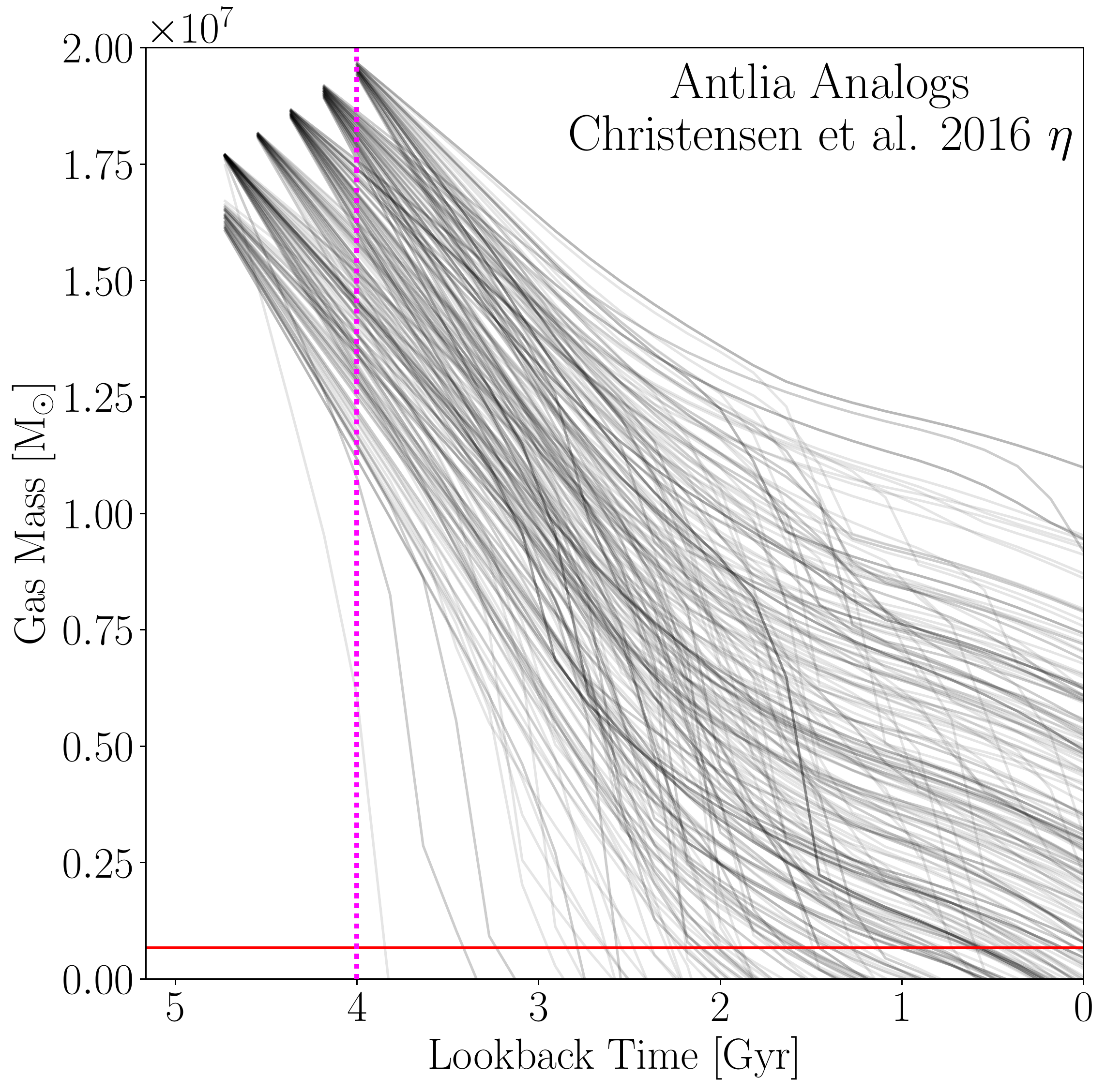}
  \caption{\emph{Left:} The total gas mass evolution including both RPS and starvation for Antlia B analogs with infall times between 3 and 4 Gyr ago. Such an infall time is relatively likely given the weighted distribution of infall times presented in Figure \ref{figure:antlia_infalls}. The weighted median infall time for all Antlia B analogs is indicated by a vertical dashed magenta line as in Figure \ref{figure:antlia_infalls}. The red horizontal line marks the observed \HI mass of Antlia B at present-day of $2.8\times10^5$ M$_{\odot}$ (see Table \ref{Table:systemproperties}). \emph{Right:} As left, but for Antlia analogs with infall times between 4--5 Gyr ago. The red horizontal line marks the observed \HI mass of Antlia at present-day of $6.8\times10^5$ M$_{\odot}$.}
\label{figure:absolute_gas_evo}
\end{figure*}

\subsubsection{Gas Mass Evolutions of Individual Systems}
Here we look at the gas mass evolutions on a per-sample basis in order to understand the variance in the quenching history of the satellite analogs. To simplify this task, we focus on samples in a narrow range of infall times identified in the previous section to give good agreement with the present-day \HI mass measurements.\par

In Figure \ref{figure:relative_gas_evo} we show the gas mass evolutions (as a fraction of infall mass) for individual Antlia and Antlia B samples with infalls roughly 3--5 Gyr ago. We choose this range of infall times because it produces analogs that agree well with the measured present-day \HI masses of Antlia and Antlia B when both starvation and RPS are active; these infall times are also highly probable given our simulation sample (Figure \ref{figure:antlia_infalls}). Most satellites in this range of infall time have experienced one to two pericenter passages. We once again separate the effects of starvation and RPS for presentational clarity. The variation in the gas mass evolutions due to starvation is a result of the dependence of the mass-loading factor on the circular velocity of the satellite halos; since the simulated analogs have different halo masses, they experience slightly different effective mass-loading factors. The variation in gas mass evolutions due to RPS is driven primarily by differences in pericenter distances and velocities. \par

Overall these plots show that there is a large degree of sample-to-sample variation in the gas mass evolutions. However, it is clear that RPS rarely removes more than $20\%$ of the initial gas mass of satellite analogs that fell into NGC 3109-like hosts between 3--5 Gyr ago, while starvation can remove almost an order of magnitude more gas. Typically $\sim12\%$ of the infall gas mass is removed by RPS, while $\sim80\%$ of the infall gas mass is removed by starvation. Only rare, highly radial infalls see greater mass loss due to RPS, and these infalls are disfavored due to the observed line-of-sight velocities and 2D projected separations of the satellites from NGC 3109. As such, the most probable orbits for Antlia and Antlia B do not result in significant mass loss due to RPS. \par

In reality, both RPS and starvation are likely to act on satellites and a holistic view of quenching should take into account all mechanisms simultaneously. We present absolute gas mass evolutions of individual analogs with both RPS and starvation active in Figure \ref{figure:absolute_gas_evo}. It is clear that the evolutions are dominated by a similar pattern of mass loss via starvation, but variations are visible due to the unique signature of RPS. \par

Due to our wide selection range of satellite halo masses (see Table \ref{Table:m_halo_constraints}) and our choice of a scaling relation for the mass-loading factor that depends on the halo circular velocity, our satellite analogs can experience quite different strengths of starvation -- we find this is actually a larger sample-to-sample variation than that introduced by different RPS strengths. \par

In summary, the quenching results for individual samples enforce our prior result based on mean evolutions that starvation is significantly more effective than RPS at quenching Antlia and Antlia B analogs in our fiducial model. \par

\subsection{Alternate Model Parameters} \label{subsection:am}

In this section we consider how varying the key parameters of our quenching models affects the results of the previous section. We identify the primary model parameters as the mass-loading factor ($\eta$) for the starvation quenching model and the host CGM density profile ($\rho_{\text{host}}(R)$) for the RPS quenching model. For the purpose of presentational clarity, we will discuss other sources of uncertainty stemming from things like initial conditions in \S \ref{subsection:uncertainties}. 

\subsubsection{Mass-Loading Factors} \label{subsubsection:massloading}
The strength of starvation as a quenching mechanism is closely tied to the  mass-loading factor. To explore how our conclusions depend on this model choice, we examine two alternate models; scaling relations from the FIRE simulations \citep{Muratov2015} and a constant $\eta=1$ as is often used in studies of the quenching of MW satellites \citep{Fillingham2015,Jahn2021,Trussler2020}. \par

The scaling relations for mass-loading factors from \cite{Muratov2015} are based on measurements from cosmological FIRE simulations \citep{Hopkins2013,Hopkins2014}. Of particular interest are two isolated dwarf galaxies in their sample with halo masses of M$_h = 2.5 \hbox{ and } 7.8 \times 10^9$ M$_{\odot}$, comparable to the expected halo masses of Antlia B and Antlia, respectively (see Table \ref{Table:m_halo_constraints}). These simulated galaxies are studied in more detail in \cite{Onorbe2015}. The  mass-loading factors reported in those works are larger than some others in the literature \citep[e.g., our fiducial model from][]{Christensen2016} but some of the difference in the normalization is due to differing outflow definitions. However, \citet{Christensen2016} find that these differing outflow definitions do not resolve the difference in the low-mass slope of the relations; \cite{Muratov2015} find $\eta \propto v_{\text{circ}}^{-3.2}$ and \citet{Christensen2016} find $\eta \propto v_{\text{circ}}^{-2.2}$. We note that the \cite{Muratov2015} relations may be superseded by \cite{Pandya2021}, who use an improved outflow definition and the most recent version of FIRE; they find the same $\eta$ scaling as \cite{Muratov2015}, but a different normalization so that their mass-loading factors are about a factor of two lower. We take the \cite{Muratov2015} values here as representative of the higher mass-loading factors in the literature. Typical mass-loading factors for Antlia and Antlia B with the \cite{Muratov2015} scaling relation are generally $40<\eta<60$ and $100<\eta<120$, respectively; these are about a factor of ten higher than the \cite{Christensen2016} values. \par

As expected, the higher mass-loading factors result in much shorter starvation quenching timescales of about 1--2 Gyr for both satellites such that starvation is even more effective than in the fiducial model. RPS removes almost no gas in comparison; only $\sim1\%$ of analogs exhibit greater than $5\%$ \HI mass loss due to RPS. If we also increase the initial gas mass by adopting the mean \cite{Papastergis2012} scaling relation between \HI mass and stellar mass, then we can obtain similar quenching timescales to the fiducial model. However, starvation remains much more effective than RPS at quenching these analogs. Coupled with our fiducial \citet{Bradford2015} gas masses (based on a larger sample of low-mass galaxies than \citealt{Papastergis2012}), the \citet{Muratov2015} mass-loading factors indicate faster quenching and more recent infall times than the infall time distributions shown in Figure \ref{figure:antlia_infalls} suggest. We show the absolute gas mass evolutions for Antlia and Antlia B analogs with infalls between 1 and 2 Gyr ago using the \cite{Muratov2015} mass-loading factors in Figure \ref{figure:c16_absolute_gas_evo}. \par

We next consider a constant $\eta=1$, corresponding to outflows equal to the instantaneous SFR; such a low mass-loading factor is often utilized in semi-analytic calculations \citep{Fillingham2015,Jahn2021,Trussler2020}. As expected, the greatly reduced mass-loading factors translate to much longer quenching timescales for the satellites, necessitating satellite infalls around 7 to 8 Gyr ago. Analogs with these infall times are able to reproduce the present-day \HI masses of Antlia and Antlia B, but such early infalls are relatively rare for the simulated analogs given their infall time distributions (Figure \ref{figure:antlia_infalls}). We additionally note that despite the earlier infall times required by this starvation model allowing for more pericenter passages, there is not a significant increase in the effectiveness of RPS. This is illustrated in the top row of Figure \ref{figure:mean_evolution} -- satellites with early infall times have lower \HI masses at infall than satellites that fall in later, and these lower \HI masses imply more centrally-concentrated gas distributions which are more resilient to RPS (see the discussion on \HI distributions in \S \ref{subsection:ic}). We show the absolute gas mass evolutions for Antlia B and Antlia analogs with infall times between 7 and 8 Gyr ago using $\eta=1$ in Figure \ref{figure:eta1_absolute_gas_evo}. \par

\begin{figure*}
  \centering
  \includegraphics[width=\textwidth,page=1]{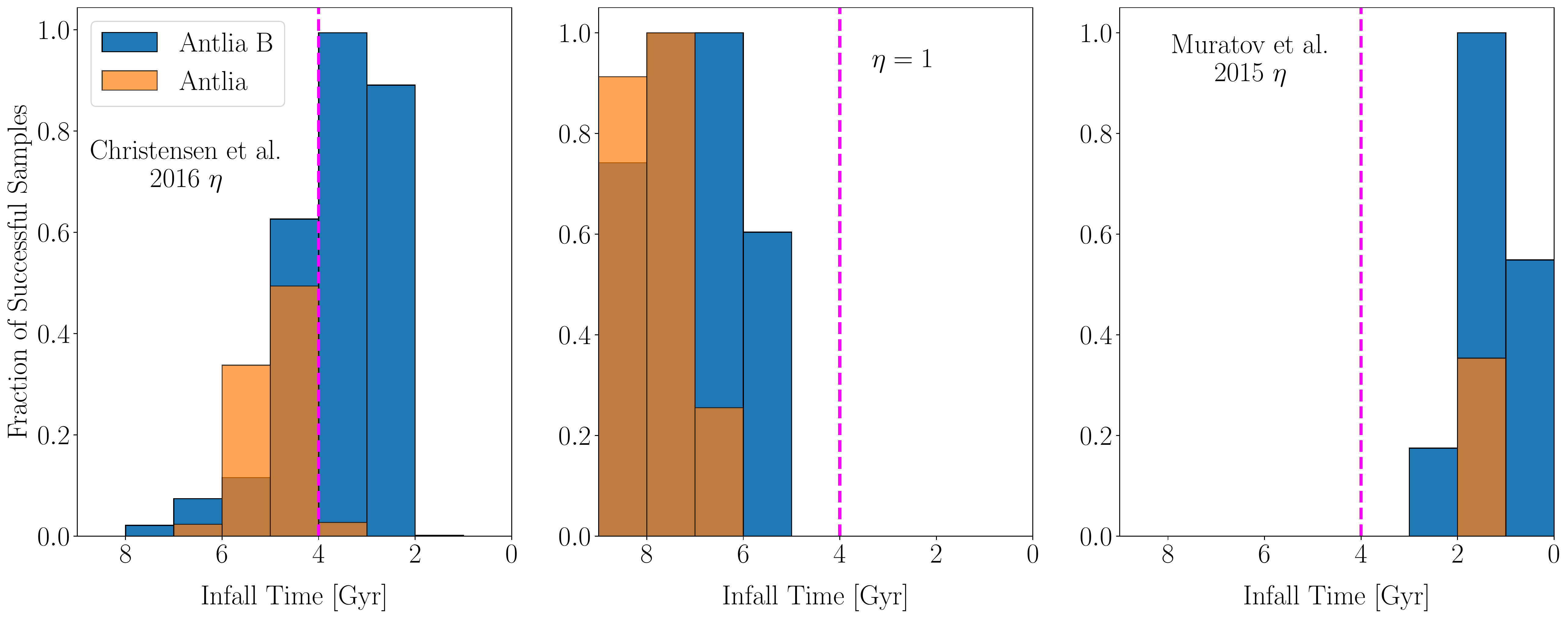} 
  \caption{Breakdown of the success of the three mass-loading factors we consider in our starvation model as a function of satellite infall time. A simulated dwarf analog is counted as a successful sample if the final \HI mass at present-day is $-2 \, \text{M}_{\text{obs},\HI} \leq \text{M}_{\HI} \leq 4 \, \text{M}_{\text{obs},\HI}$. The weighted median infall time of Antlia and Antlia B analogs is $\sim4$ Gyr, indicated by vertical dashed magenta lines. It is clear that the fiducial \protect\cite{Christensen2016} mass-loading factors give the best agreement with the median infall times, while $\eta=1$ requires earlier infalls to reproduce the observed \HI masses, and the \protect\cite{Muratov2015} mass-loading factors require later infalls. }
\label{figure:percentsuccess}
\end{figure*}

To summarize the results from our different mass-loading factor models, we show a breakdown of the success of the models as a function of satellite infall time in Figure \ref{figure:percentsuccess}. A simulated analog is a ``successful" sample if its final \HI mass at present-day is $-2 \, \text{M}_{\text{obs},\HI} \leq \text{M}_{\HI} \leq 4 \, \text{M}_{\text{obs},\HI}$, with this range chosen to highlight the differences between the models. It is clear from this comparison that the fiducial \cite{Christensen2016} mass-loading factors are completely consistent with the weighted median infall times of Antlia and Antlia B, which are both $\sim4$ Gyr (see Figure \ref{figure:antlia_infalls}). With $\eta=1$, quenching is much slower, necessitating considerably earlier infalls, which are rare for our simulated samples. Conversely, the high \cite{Muratov2015} mass-loading factors require quite late infall times, generally within the last 2 Gyr. This demonstrates the importance of including realistic mass-loading factors when considering the quenching timescales of low-mass dwarfs post-infall. \par

\subsubsection{Alternate Host CGM Models} \label{subsubsection:hostcgm}
As the ram pressure experienced by the satellite is linearly related to the host CGM density (Equation \ref{equation:rpsbasic}), our conclusions about the efficacy of RPS depend directly on our assumptions about the host CGM. Though our fiducial singular isothermal sphere density model is well-supported by simulations \citep[e.g.,][]{Fielding2017,Hafen2019a}, the normalization (and thus, the total mass in the CGM) is significantly more uncertain, especially for galaxies as low-mass as NGC 3109. Under our fiducial normalization, the total mass in the CGM of NGC 3109 within 100 kpc of its center is $\sim10^9$ M$_\odot$; here we examine changes in the effectiveness of RPS when increasing the total CGM mass of our hosts.\par

Under our fiducial host CGM model, Antlia and Antlia B analogs with infalls around 4 Gyr ago lose $\sim10\%$ of their infall \HI mass due to RPS (see the left column of Figure \ref{figure:relative_gas_evo}). If we increase the host CGM mass by a factor of two, to $\sim2\times10^9$ M$_\odot$, we see an additional $\sim4\%$ mass loss due to RPS, for an average \HI mass loss due to RPS of $\sim14\%$ of the infall \HI mass. Analogs with earlier infall times still do not experience significant mass loss due to RPS because their \HI distributions are more compact at infall (see the top row of Figure \ref{figure:mean_evolution}). If we increase the host CGM mass by a factor of four, to $\sim4\times10^9$ M$_\odot$, we see a greater change in the \HI mass loss, which becomes $\sim22\%$ of the infall \HI mass. The CGM mass must be $\sim2.5\times10^{10}$ M$_\odot$ before we reach $50\%$ mass loss due to RPS, a factor of 25 greater than our fiducial value and nearly equal to our estimated halo mass for NGC 3109 of $3.87\times10^{10}$ M$_\odot$ (see Table \ref{Table:m_halo_constraints}). \par

Even with a significantly increased CGM mass, we do not find it plausible that Antlia and Antlia B were quenched via RPS. Even with a host CGM four times more massive than in our fiducial model, Antlia and Antlia B analogs with infalls around 4 Gyr ago are too \HI-rich at present day, having 5--10 times more mass in \HI than is measured for the real satellites. Analogs with earlier infall times are no better; they experience less mass loss due to RPS because they have more compact \HI distributions at infall. This reinforces our conclusion that starvation is necessary for the quenching of Antlia and Antlia B and shows that our results are robust to uncertainties in the host CGM normalization.\par

\subsection{Model Uncertainties} \label{subsection:uncertainties}
Having examined the dependence of our results on the key model parameters, we now move on to discuss other sources of uncertainty in our analysis. These relate in particular to initial conditions (e.g., the initial \HI masses of the satellites at infall) and observational quantities of the NGC 3109 system (e.g., the assumed SFHs of the satellites). 

\subsubsection{Initial \HI Mass} \label{subsubsection:himass}
Our experimental setup with a fixed, empirical SFH is quite sensitive to the choice of satellite \HI mass at infall. We demonstrated this in \S\ref{subsubsection:massloading}, where we computed gas mass evolutions under the \cite{Muratov2015} mass-loading factors with two different models for satellite \HI masses at infall. For these high mass-loading factors, the change in infall \HI masses led to satellite quenching times that were a factor of $\sim2$ different. However, we have thusfar neglected scatter in the empirical relation we assume for our fiducial infall \HI masses \citep{Bradford2015}. Given the dependence of the quenching behavior on the initial \HI mass, it is reasonable to question what effect including the empirical scatter in our model has on our conclusions.\par 

With the simplicity of the quenching models, we can reason about the effects of including uncertainty on the initial \HI. With our fiducial model, the initial \HI masses of Antlia and Antlia B analogs are typically $\sim25$ and $\sim5$ times greater than their present-day measured \HI masses, respectively. If we model the intrinsic scatter as lognormal, as indicated in the literature \citep[e.g.,][]{Papastergis2012,Bradford2015,Scoville2017}, it is clear that the spread in the initial \HI masses will be comparable to or greater in magnitude than the target final \HI mass. For instance, if the expected initial \HI mass at a given infall time for an Antlia B analog is $1.5\times10^6$ M$_\odot$, then a 0.2 dex lognormal scatter results in a 1-$\sigma$ range of roughly $1.5^{+0.8}_{-0.5}\times10^6$ M$_\odot$ at fixed infall time, while the present-day observational value is $3\times10^5$ M$_\odot$. \par

\begin{figure}
  \centering
  \includegraphics[width=0.45\textwidth,page=1]{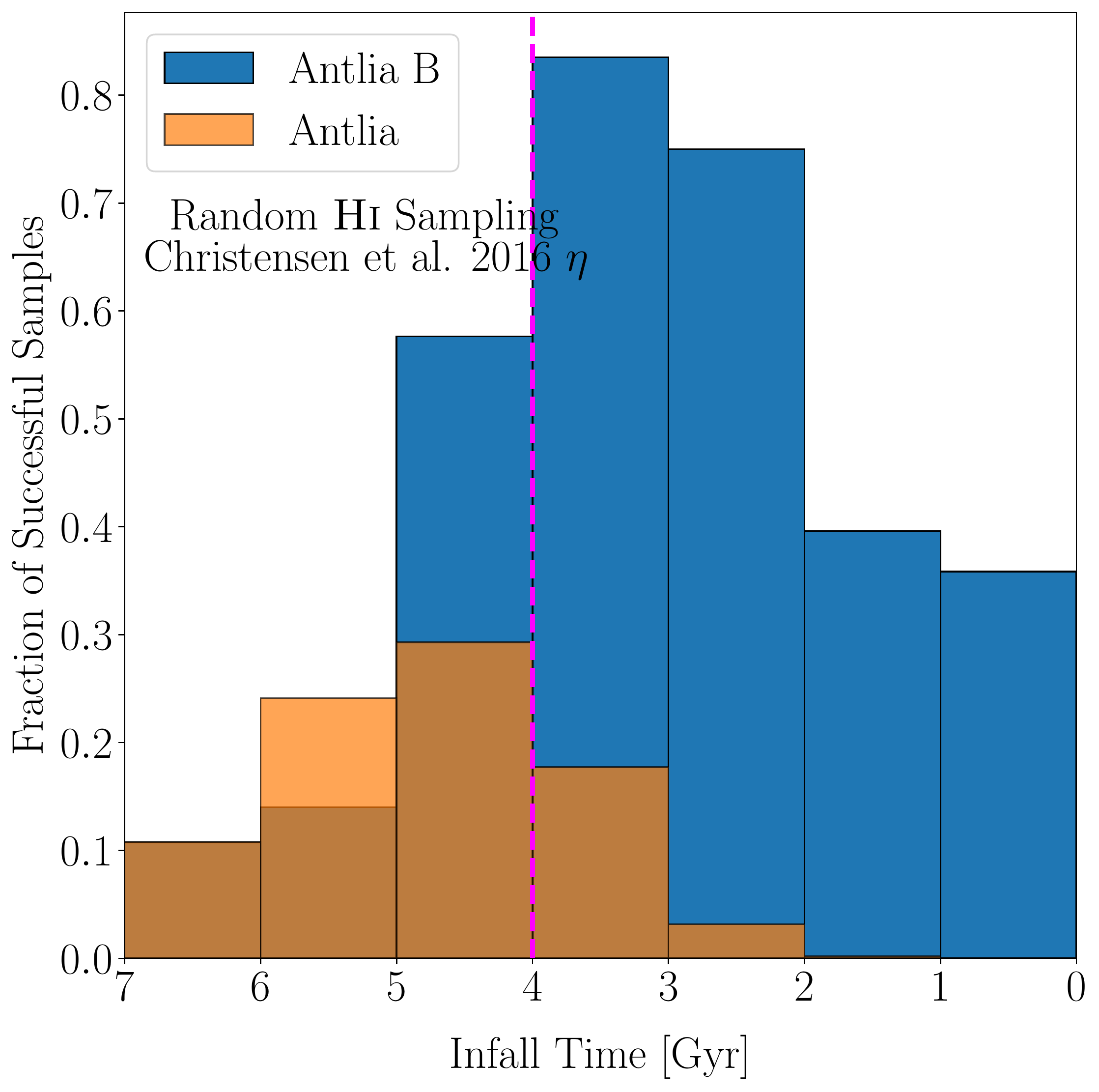}
  \caption{The fraction of analog satellites in 1 Gyr bins of infall time with final \HI masses at present-day between $-2 \, \text{M}_{\text{obs},\HI} \leq \text{M}_{\HI} \leq 4 \, \text{M}_{\text{obs},\HI}$ when randomly sampling the \HI mass of the satellites at infall according to the empirical scatter \citep{Bradford2015}. This is analogous to the left panel of Figure \ref{figure:percentsuccess}, but includes the empirical uncertainty on the \HI masses at infall. In comparison, including uncertainty on the infall \HI masses broadens the distribution of plausible infall times, but does not meaningfully change our conclusions.}
\label{figure:rand_hi}
\end{figure}

We can see how this uncertainty affects our analog samples in Figure \ref{figure:rand_hi}, which is analogous to the left panel of Figure \ref{figure:percentsuccess}, but where we have randomly sampled the initial \HI masses of the satellites at infall according to a lognormal distribution with 0.2 dex of scatter as discussed above. Including this uncertainty allows for a broader range of analog infall times to produce final \HI masses that are comparable to the observed values for the satellites. This effect is stronger for more recent infall times, as more recent infalls have higher mean \HI masses at infall under our fiducial model, leading to larger scatter under the lognormal distribution. Including this uncertainty does not lead to any significant increase in the average effect of RPS; it remains relatively ineffective at removing gas from our satellite analogs in the presence of this uncertainty. We additionally see that when accounting for this uncertainty, the mean infall times of our simulated analogs still agree well with the distribution of successful samples. \par

In addition to this large intrinsic scatter, there is also some disagreement in the literature about the low-mass slope of the M$_*$--M$_\HI$ relation; if we adopt instead the power-law fit of \cite{Papastergis2012}, we find expected initial \HI masses for Antlia and Antlia B that are factors of 2.5 and 7.5 higher than the \cite{Bradford2015} values. As we have shown that the simulated present-day \HI masses scale almost linearly with the initial \HI masses, adopting the \cite{Papastergis2012} relation would translate to quenching timescales a few Gyr longer than we obtain with the fiducial \HI masses based on the results of \cite{Bradford2015}. The larger initial \HI masses result in more extended satellite \HI distributions under our surface density model (\S\ref{subsection:ic}) such that RPS is slightly more effective, stripping on average $\sim15-20\%$ of the infall \HI masses of Antlia and Antlia B analogs compared to an average of $\sim10\%$ under our fiducial model. However, significantly more gas needs to be removed to reproduce the present-day satellite \HI masses, and RPS cannot remove all of this gas on its own -- the majority of analogs end up with present-day \HI masses that are at least 5--10 times higher than the observed values under the effects of RPS alone. \par 

As the large uncertainty in the initial \HI mass propagates to the final present-day \HI mass, the quenching timescales quoted throughout the paper should be regarded as expectation values with relatively large uncertainties. However, our conclusion that starvation is more effective at quenching Antlia and Antlia B analogs is robust in the presence of this uncertainty due to the weak scaling of RPS mass loss with the initial \HI mass for our adopted gas surface density model. \par

\subsubsection{Star Formation Histories and Stellar Masses}
In this section, we consider the effects of the uncertainty in the SFHs of the two satellites on our inferences about the relative efficacy of quenching mechanisms.  For SFHs derived from resolved stellar photometry, such as we use here, there is generally an anti-correlation between adjacent time bins in the absolute SFH (i.e., the total amount of stellar mass formed in a time bin) because if the SFR is overestimated in bin $i$, it is typically underestimated in bin $i+1$ due to the similarities in the color-magnitude diagrams \citep{Weisz2011,Dolphin2013}. However, this trait is removed by constructing a cumulative SFH (i.e., the sum of all stellar mass formed up to and including bin $i$) and normalizing it to the final integrated stellar mass. In a cumulative SFH, the uncertainties in adjacent time bins are uncorrelated. We use these cumulative SFHs for our analysis and normalize them to the stellar masses in Table \ref{Table:systemproperties}. In our simple starvation model, neglecting time-evolution of the mass-loading factor $\eta$, the present-day gas mass considering only starvation is simply $\text{M}_{\HI,z=0} = \text{M}_{\HI,\text{infall}} + (R - 1 - \eta) \, \Delta \text{M}_*$ where $\Delta\text{M}_*=\text{M}_{*,z=0} - \text{M}_{*,\text{infall}}$ is the total stellar mass formed between infall and the present-day. Since $\eta$ is at least an order of magnitude greater than $R$ for the dwarf masses considered here, the scatter in the present-day \HI mass will approximately scale with $\eta$ and the uncertainty in $\text{M}_{*,z=0}$. \par

As the uncertainties in the present-day stellar masses are of the same order as the present-day \HI masses, and the time-averaged $\eta$ for Antlia and Antlia B are about 6 and 10, respectively, the stellar mass uncertainties introduce a large uncertainty on the present-day \HI mass. This uncertainty is subdominant to the uncertainty in the initial \HI mass for Antlia analogs, but is about a factor of 2--4 greater than the uncertainty in the initial \HI mass for Antlia B analogs due to the larger mass-loading factors of Antlia B analogs. It should also be clear that since the initial \HI masses are based on the stellar masses, uncertainty in the stellar masses will also increase the scatter in the initial \HI masses. We do not explore this correlation further, as we have demonstrated that the uncertainties in the initial \HI masses and SFHs introduce significant errors into the present-day \HI mass estimates at fixed infall time. We reiterate that, given these large uncertainties, the absolute quenching timescales derived in \S\ref{subsection:fiducialmodel} are largely uncertain, but conclusions about the relative efficacy of starvation and RPS remain robust. \par


\subsubsection{Tidal Stripping} \label{subsubsection:tidalstripping}
In our fiducial model we neglect tidal stripping in order to simplify the orbital evolution. However, it is worth considering whether tidal stripping affects the gas reservoirs of our analog satellites. If the tidal radius is at any point smaller than the historical minimum of the RPS stripping radius, then tidal stripping may remove additional gas from the satellite. Adopting the definition of tidal radius from \cite{Penarrubia2008} of $\langle \rho_{\text{sat}}(r_t)\rangle = 3 \, \langle \rho_{\text{host}}(R)\rangle$ where $\langle \rho_{\text{sat}}(r_t)\rangle$ is the mean dark matter density of the satellite inside the tidal radius and $\langle \rho_{\text{host}}(R)\rangle$ is the mean dark matter density of the host inside $R$, which is the distance between the host and satellite centers. We observe median pericenters of $\sim 35$ and $\sim45$ kpc for Antlia and Antlia B analogs, respectively, and find tidal stripping radii at these pericenters greater than 10 kpc, at least twice as large as the RPS stripping radius. The tidal radius is similar to the RPS stripping radius for pericenters around 10 kpc, with a typical value of $\sim3$ kpc for Antlia analogs and $\sim1$ kpc for Antlia B analogs, though we have few analog samples with such small pericenters. For pericenters closer than 10 kpc, tidal stripping can remove gas if the velocity of the satellite is low enough, but only $\sim7\%$ of the analog satellites have such close pericenters (Figure \ref{figure:antlias_orbit}). It therefore seems unlikely that tidal stripping is a significant quenching mechanism for hosts of NGC 3109's mass. \par

Given these calculations, we can comment briefly on the observation of elongation of the stellar component of Antlia presented in \cite{Penny2012}, which they suggested may be due to tidal disruption. Given Antlia's stellar effective radius of $\sim500$ pc and assuming that the full extent of the stellar population is $\sim1$ kpc, we see that the tidal stripping radius is beyond the stellar radius for even very close pericenters ($<10$ kpc), due in large part to the fact that our expected halo mass for Antlia is roughly a quarter that of NGC 3109 and so is larger than assumed in their work. As such, for our simulation sample and adopted halo mass ranges, tidal stripping of Antlia's stars seems quite unlikely. However, it is still possible, as they argue, that an increase in the internal binding energy of the satellite may lead to partial dissolution. A full exploration of the possible tidal disruption of Antlia's stellar population is beyond the scope of this work. \par

\section{Conclusion} \label{section:conclusion}
In order to study the quenching mechanisms relevant for satellites of low-mass hosts, we have selected systems analogous to NGC 3109, which hosts two satellites, Antlia and Antlia B, both of which have well-measured \HI masses and SFHs (see Table \ref{Table:systemproperties} for observational properties and Table \ref{Table:m_halo_constraints} for halo mass estimates). Analogs are selected from the cosmological TNG100 simulation with hydrodynamics \citep{Nelson2018,Naiman2018,Springel2018,Pillepich2018a,Marinacci2018}. We derive halo mass probability distributions for the observed galaxies, allowing us to select a larger sample of analogs and propagate the probability that they represent the observed system through our analysis. We additionally derive probability distributions for the projected separations and line-of-sight velocities of the simulated systems and utilize the observational data to further constrain our analog sample. \par

With simulated analogs in hand, we construct an observationally-constrained semi-analytic model to study the evolution of the gas masses of the satellites after infall. We implement gas mass loss due RPS and starvation (i.e., cessation of cold gas inflows) in the semi-analytic model, and examine tidal stripping in post-processing. Because we estimate that Antlia was about $25\%$ as massive as NGC 3109 at first infall, we resimulate the orbits of all systems, including a model for dynamical friction, from first infall to the present-day to properly model RPS. Rather than self-consistently evolving the star formation, we fix the SFHs of the satellites to the observed values when computing the mass loss due to starvation.\par

For our fiducial quenching models, we find that starvation is much more effective than RPS. In particular, $\sim95\%$ of analog satellites have less than $20\%$ of their initial \HI removed by RPS, which is insufficient to produce the present-day observed \HI masses unless the infall gas masses of the satellites were an order of magnitude lower than our fiducial values. Only for rare ($<1\%$ of samples), highly radial orbits is RPS able to strip a significant fraction of gas. We additionally show that the tidal stripping radius is almost always larger than the RPS stripping radius in our model, such that tidal stripping is incapable of removing a significant amount of gas under the vast majority of likely satellite orbits. In contrast, we find starvation to be highly effective, producing reasonable agreement with the observed present-day gas masses for infall times between 3 and 4 Gyr ago for Antlia and 4 to 5 Gyr ago for Antlia B, squarely in the middle of the infall time probability distributions indicated in the TNG100 simulations. While absolute quenching timescales are difficult to constrain due to uncertainties in the initial conditions for our semi-analytic model, it is clear from our results that starvation is the primary quenching mechanism for Antlia and Antlia B under our fiducial model choices. \par

To survey the range of model parameters supported by the literature, we examine two alternate mass-loading factors, including a constant $\eta=1$ and the relation between $\eta$ and M$_*$ given in \cite{Muratov2015}, with the former being lower than the fiducial values from \cite{Christensen2016} and the latter being higher. For the \cite{Muratov2015} relation, we find much shorter quenching timescales as expected, with starvation being perhaps \emph{too} effective, as the quenching timescales for starvation with the \cite{Muratov2015} model are as short as 1--2 Gyr. For $\eta=1$, we find much longer quenching timescales that imply first infalls around 7 to 8 Gyr ago, which are unlikely for our sample; importantly, we still find that starvation is the main mechanism that removes gas from the dwarfs, even with such a low mass-loading factor. \par

Our results suggest that starvation is significantly more effective than RPS at removing gas from the satellites of such low-mass hosts. This is in contrast to some recent work with hydrodynamic simulations \citep[e.g.,][]{Jahn2021} that suggests RPS is the dominant quenching mechanism at this mass scale. However, it is difficult to differentiate between starvation and RPS in hydrodynamic simulations because energetic stellar feedback may ``loosen" the gas of the dwarf satellites, allowing it to be more easily stripped by RPS. Such ``weak" RPS, which primarily only removes such loosened gas, could be differentiated from ``strong" RPS that directly strips the dense interstellar media of the dwarf satellites (as implemented in our model) by testing whether gas particles ejected from the dwarf galaxy were recently exposed to stellar feedback (e.g., a supernova).\par 

Given these simulation results, we find it likely that ``weak" RPS may explain starvation in low-mass systems by both intercepting pristine gas inflows and preventing satellites from re-accreting metal-enriched outflows. Such a scenario does not require ``strong" RPS to directly remove cold, dense \HI from the disks of accreted satellites, which we have shown to be ineffective at these mass scales. This scenario does, however, require that low-mass galaxies have circumgalactic media or bulk outflows that are capable of causing this ``weak" RPS as is indicated by recent work at mass scales slightly above that of NGC 3109 \citep[e.g.,][]{Bordoloi2014,Johnson2017,Fielding2017,Hafen2019a,Jahn2021,Pandya2021}. Given these results, our work with the satellites of the NGC 3109 system suggests that outflows play a major dual role in the evolution of satellites: outflows from low-mass hosts cause ``starvation" by preventing gas from being accreted onto dwarf satellites, and outflows from dwarf satellites dramatically shorten the timescale for quenching once satellites are disconnected from their gas supply. Extending observational and theoretical studies of star-formation-driven outflows and galactic CGM to lower galaxy masses will shed further light on this subject.

\hspace{\linewidth}
\section*{Acknowledgements}
We would like to thank Stephanie Tonnesen for helpful discussions regarding ram pressure stripping. We also thank Carton Zeng for helpful comments. C.T.G. and A.H.G.P. are primarily supported by the National Science Foundation (NSF) Grant No. AST-1813628, and additionally supported by NSF Grant No. AST-1615838. K.S. acknowledges support from the Natural Sciences and Engineering Research Council of Canada (NSERC). D.J.S. acknowledges support from NSF grants AST-1821967 and 1813708. Research by D.C. is supported by NSF grant AST-1814208. J.L.C. acknowledges support from HST grant HST-GO-15228 and NSF grant AST-1816196.
\par
This research has made use of the NASA/IPAC Extragalactic Database (NED), which is operated by the Jet Propulsion Laboratory, California Institute of Technology, under contract with the National Aeronautics and Space Administration. This research has made use of NASA's Astrophysics Data System.

\bibliographystyle{mnras}
\bibliography{library}

\appendix
\section{Utilizing Projected Quantities} \label{appendix:pq}
In the past it has been common to use projected quantities of the observed system, such as the 2D separation and line-of-sight velocity difference, to further constrain the set of analogs \citep[e.g.,][]{Sales2013,Besla2018,Garling2020}. This is typically implemented with Monte Carlo rejection sampling, where random rotation matrices are sampled and applied to the 3D positions and velocities of simulated subhalos before calculating projected quantities along a chosen axis of observation. If these projected quantities match some set selection range, the subhalo is saved for later analysis. This procedure produces a better set of subhalo analogs because the projected quantities of the system we observe are correlated with their 3D quantities and thus contain useful information about the system. However, rejection sampling is a suboptimal method for this calculation. It is computationally inefficient, as a large portion of the samples are discarded on each iteration, and it is prone to numerical error when the sample size itself is small, as may be the case when working with small simulation volumes or very specific analog selections. To avoid these issues, we derive directly the probabilities of subhalos with given 3D positions and velocities having specific projected quantities. We first consider the projected separation. \par

Random observation of a host-subhalo system is equivalent to stating that the host-centric coordinate axes are randomly aligned with respect to our observation point; i.e., in spherical coordinates, the $\phi$ (polar angle) and $\theta$ (azimuthal angle) coordinates of the subhalo are uniformly distributed over the sphere. However, the host-centric distance of the subhalo is fixed ($r$ in spherical coordinates). If we choose to observe along the Cartesian $z$ axis, we find a joint PDF of 

\begin{equation}
  \begin{aligned}
    & \frac{dP (x,\phi,\theta|r)}{dx \, d\phi \, d\theta} =\frac{1}{4\pi} \ \sin{\phi} \\
    &\delta\left(r \ \sqrt{\left[\sin{\phi}\cos{\theta}\right]^2 + \left[\sin{\phi}\sin{\theta}\right]^2} - x\right) 
    \end{aligned}
\end{equation}

\noindent where $\delta(X)$ is the Dirac delta function and $x$ is the projected separation between the host and satellite. Marginalization over $\phi$ and $\theta$ gives the PDF for $x$ given $r$,

\begin{equation} \label{equation:proj_prob}
  \frac{dP (x|r)}{dx} =\frac{x \ \heaviside{r-x}}{r^2 \ \sqrt{1-\frac{x^2}{r^2}}}
\end{equation}

\noindent where $\heaviside{X}$ is the Heaviside function, which is 0 when $X<0$ and 1 when $X>0$. The PDF diverges as $\lim_{x \to r} \ \frac{dP \, (x|r)}{dx} =\infty$, though its integrated probability is finite with $\int_0^r \, \frac{dP(x|r)}{dx} \, dx = 1$. Given a PDF for the observed value of $x$ given the data $\mathcal{D}$, denoted $\frac{dP(x|\mathcal{D})}{dx}$, the probability that a random observation of a subhalo with host-centric distance $r$ is consistent with the observation is

\begin{equation} \label{equation:proj_prob2}
  P(x|r) =\int_0^r \frac{x}{r^2 \ \sqrt{1-\frac{x^2}{r^2}}} \, \frac{dP(x|\mathcal{D})}{dx} \, dx
\end{equation}

\noindent Though we find no position uncertainties for NGC 3109 in the literature, determining the center of such a dwarf irregular is difficult and cannot be done to infinite precision. We adopt a Gaussian distribution for $\frac{dP(x|\mathcal{D})}{dx}$ with $\sigma=1$ kpc as an upper limit to the uncertainty. \par

We next consider the line-of-sight velocity difference ($\Delta v$) between satellite and the host. The velocity problem can be thought of analogously to the position problem; the magnitude of a subhalo's 3D velocity difference from its host is fixed, but the orientation of the velocity vector with respect to our observational axis is random. For a subhalo with a velocity difference vector $\mathbf{V}=\mathbf{V}_{\text{sat}}-\mathbf{V}_{\text{host}}$ with norm $|\mathbf{V}|=V$, the probability of measuring a line-of-sight velocity $\Delta v$ is uniform from $-V$ to $V$, such that $\frac{dP(\Delta v|V)}{d\Delta v}=\heaviside{V-|\Delta v|}/2V$. If the data, $\mathcal{D}$, include some uncertainty on the observational line-of-sight velocity difference described by the PDF $\frac{dP(\Delta v|\mathcal{D})}{d\Delta v}$, we can write the probability that a random observation of $\mathbf{V}$ will match the observed value as

\begin{equation}
    P(\Delta v | V) = \frac{1}{2V} \ \int_{-V}^{V} \frac{dP(\Delta v|\mathcal{D})}{d\Delta v} \, d \Delta v
\end{equation}

\noindent If $\frac{dP(\Delta v|\mathcal{D})}{d\Delta v}$ is Gaussian with mean $\mu$ and standard deviation $\sigma$, then

\begin{equation}
  \begin{aligned}
    P(\Delta v | V) &= \frac{1}{4V} \left[ \erf{\frac{V - \mu}{\sqrt{2} \ \sigma}} + \erf{\frac{V+\mu}{\sqrt{2} \ \sigma}} \right]
  \end{aligned}
\end{equation}

\noindent We take the measured line-of-sight velocities and uncertainties from Table \ref{Table:systemproperties} for $\mu$ and $\sigma$. The probabilities are then multiplied into the weights from Equation \ref{equation:simprob} as

\begin{equation}
  \begin{aligned} \label{equation:simprob2}
    w_i=&P(\text{M}_{h,\text{sat},i}|\text{M}_{*,\text{sat}}) \ P(\text{M}_{h,\text{host},i}|\text{M}_{*,\text{host}}) \\
    & P(x|r) \ P(\Delta v| V )
  \end{aligned}
\end{equation}


\section{Gas Mass Evolutions for Alternate Mass-Loading Factors}
\begin{figure*}
  \centering
  \includegraphics[width=0.45\textwidth,page=1]{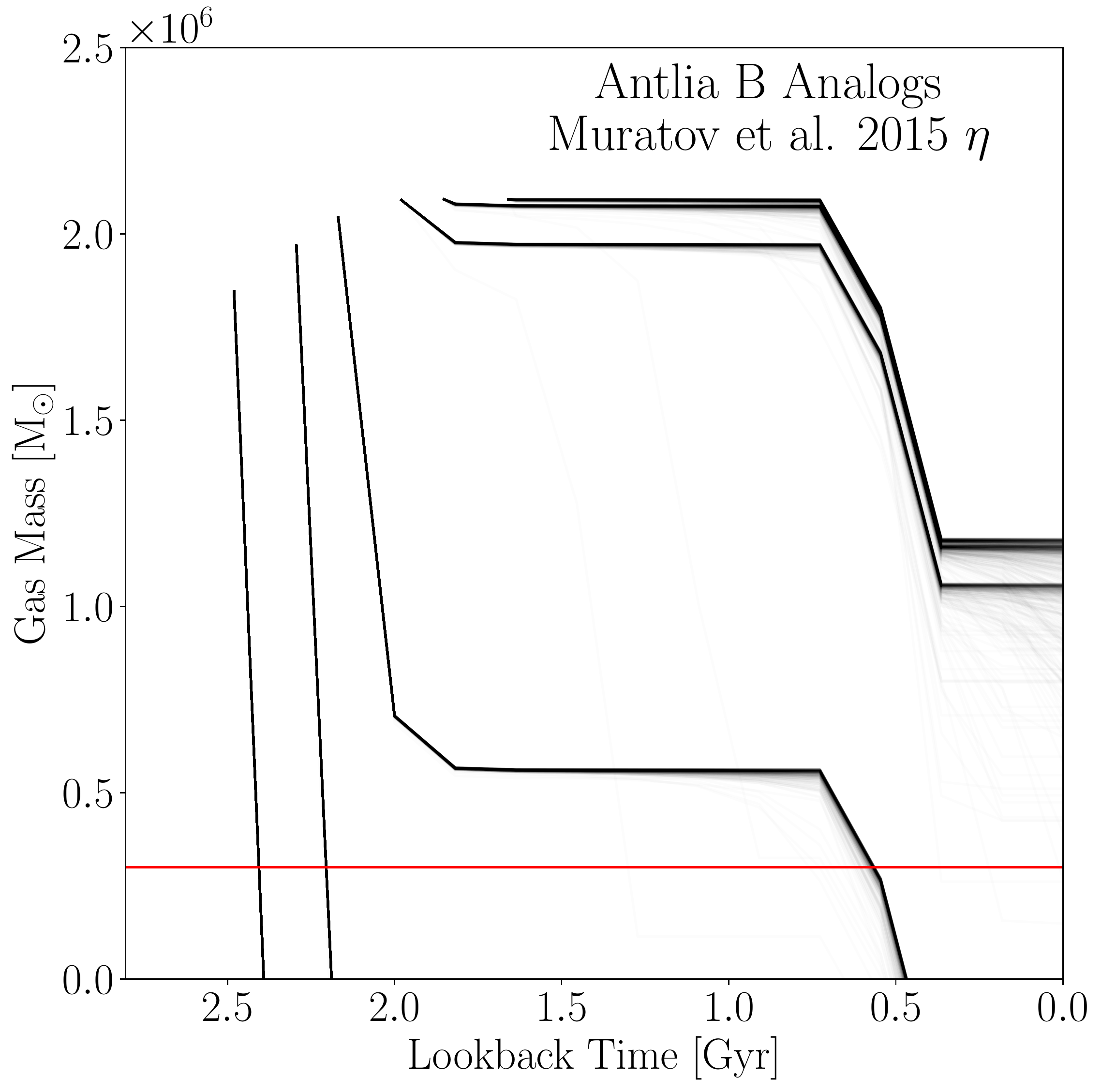} \hspace{1cm}
  \includegraphics[width=0.45\textwidth,page=1]{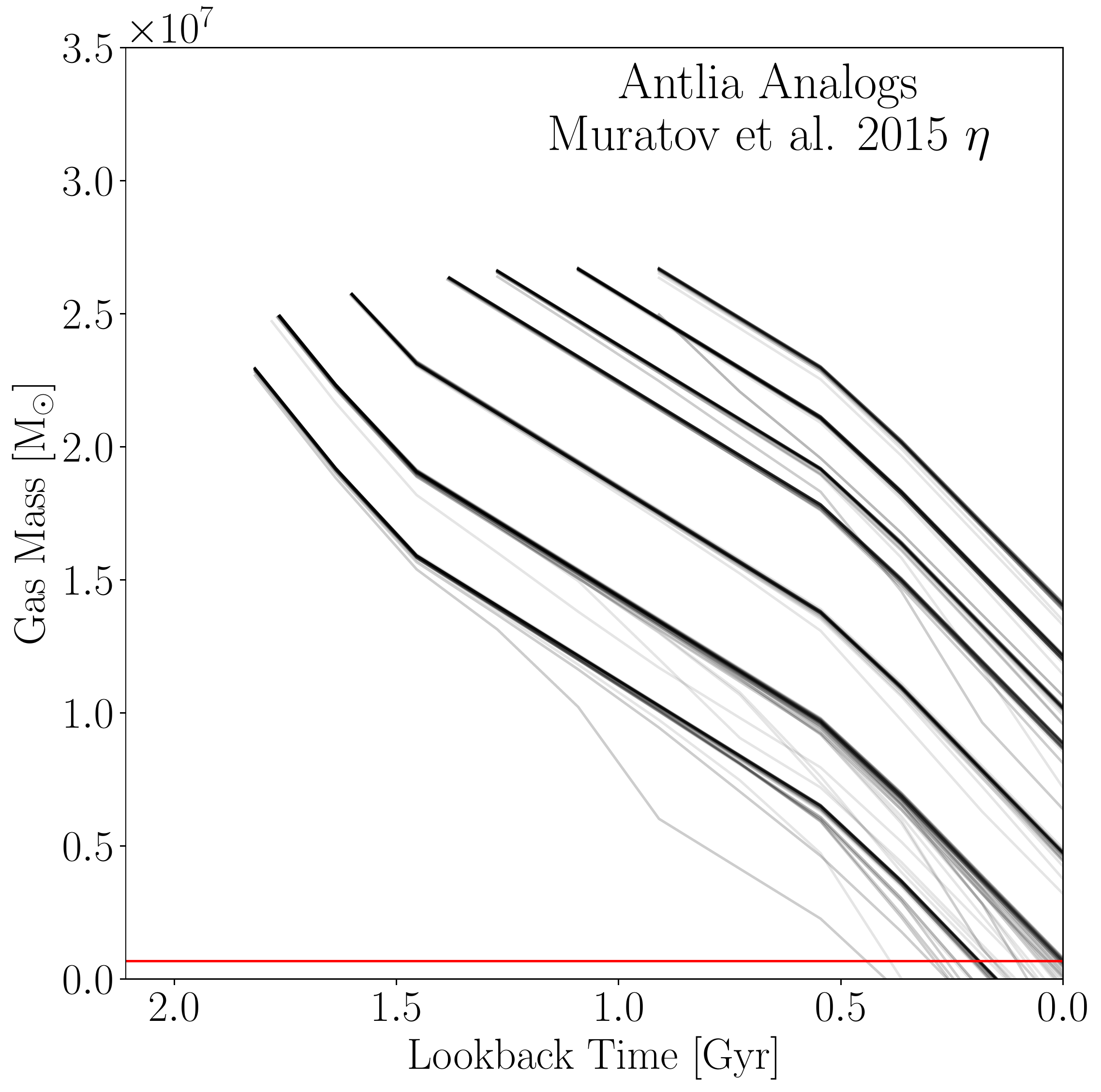}
  \caption{The total gas mass evolution including RPS and starvation for Antlia B (left) and Antlia (right) analogs with infall times between 1 and 2 Gyr ago computed with the \protect\cite{Muratov2015} mass-loading factors. These mass-loading factors are about a factor of ten higher than the fiducial \protect\cite{Christensen2016} mass-loading factors, leading to much shorter quenching timescales due to starvation. The weighted median infall times for the Antlia and Antlia B analog samples are both about 4 Gyr, considerably earlier than required by this outflow model. The red horizontal lines mark the observed \HI masses of Antlia and Antlia B, respectively (see Table \protect\ref{Table:systemproperties}).}
\label{figure:c16_absolute_gas_evo}
\end{figure*}

\begin{figure*}
  \centering
  \includegraphics[width=0.45\textwidth,page=1]{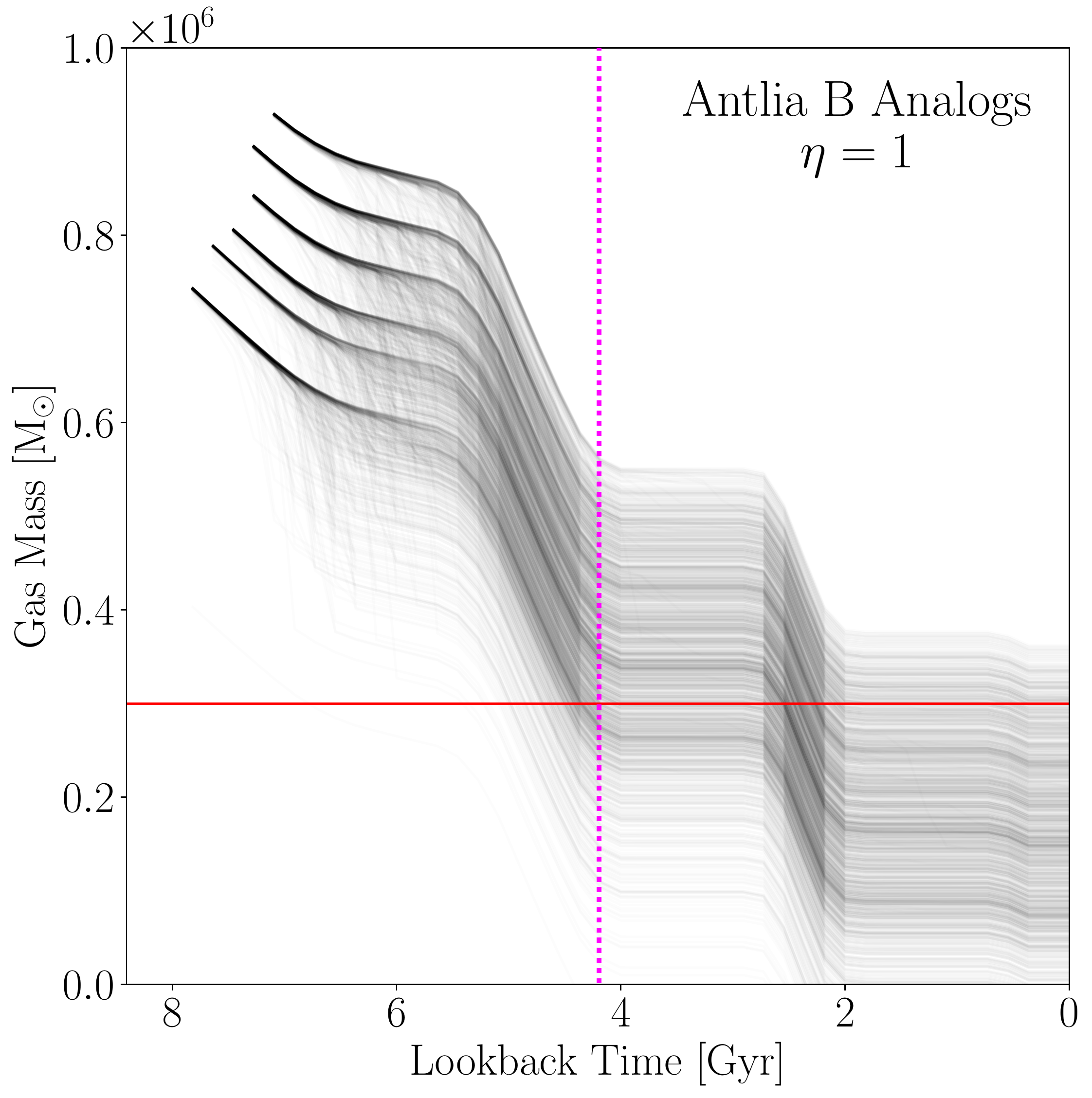} \hspace{1cm}
  \includegraphics[width=0.45\textwidth,page=1]{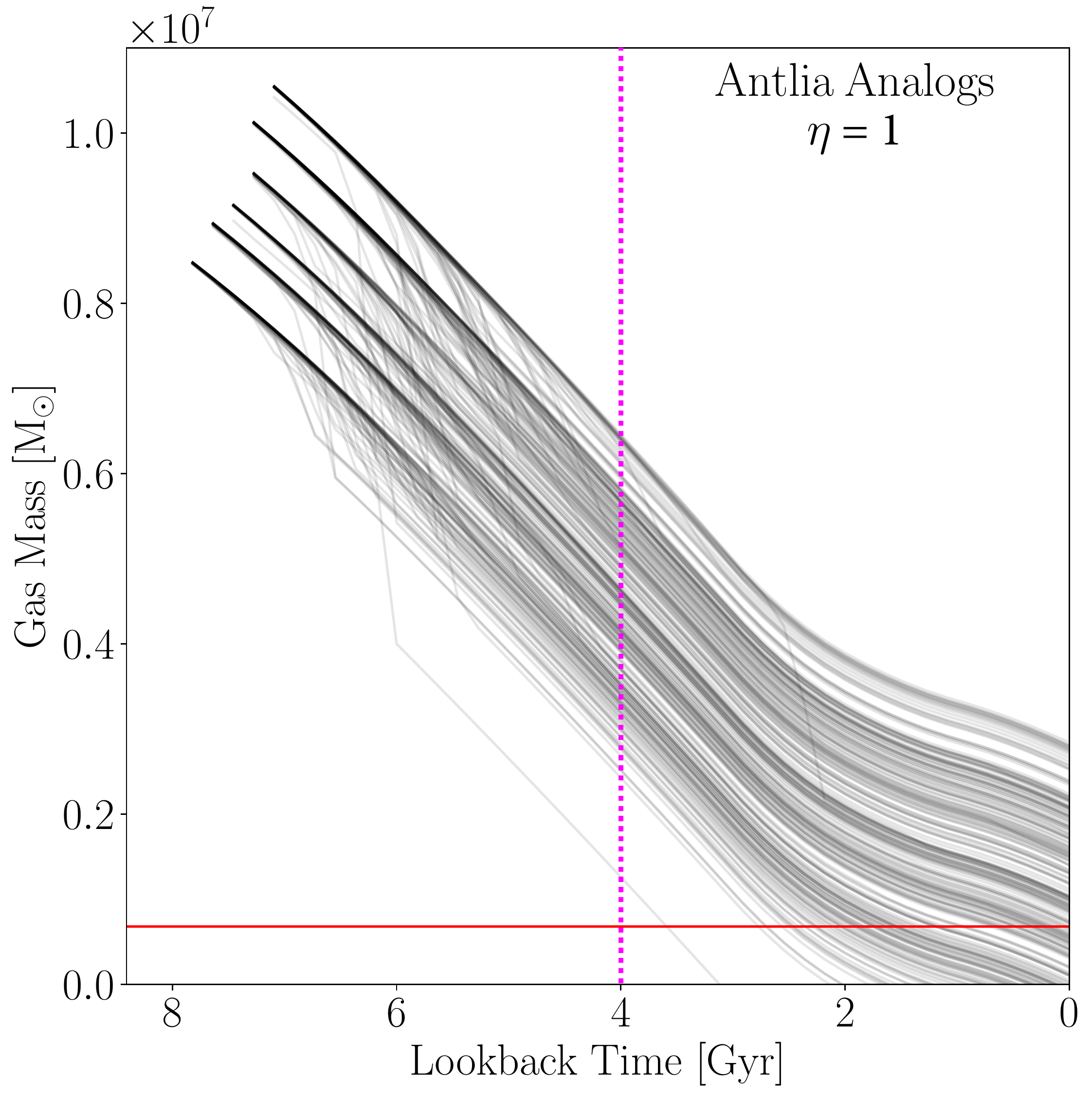}
  \caption{The total gas mass evolution including RPS and starvation for Antlia B (left) and Antlia (right) analogs with infall times between 7 and 8 Gyr ago computed with constant mass-loading factors of 1. These infall time bins were chosen to produce present-day satellite gas masses that are consistent with the observations. The red horizontal lines mark the observed present-day \HI masses of Antlia and Antlia B, respectively (see Table \ref{Table:systemproperties}). A mass-loading factor of unity is sometimes used in semi-analytic calculations, but requires very early infall times for Antlia and Antlia B; the weighted median infall times for the analog satellite populations are indicated by vertical magenta dashed lines as in Figure \ref{figure:antlia_infalls}.}
\label{figure:eta1_absolute_gas_evo}
\end{figure*}

We present here absolute gas mass evolutions for our simulated Antlia and Antlia B analogs (analogous to Figure \ref{figure:absolute_gas_evo}) under the effects of both RPS and starvation, but with alternate models for the mass-loading factors. In Figure \ref{figure:c16_absolute_gas_evo} we show the absolute gas mass evolutions for Antlia B and Antlia analogs using the mass-loading factors from \cite{Muratov2015} which are about a factor of ten greater than the fiducial \cite{Christensen2016} values. This leads to significantly reduced quenching times, necessitating more recent infalls (generally 1--2 Gyr ago) for the satellites in order to reproduce their present-day measured \HI masses. In contrast, Figure \ref{figure:eta1_absolute_gas_evo} shows analogous results but for a constant mass-loading factor of 1, as is sometimes used in semi-analytic calculations. In this case, the quenching timescales are much longer than in the fiducial case, necessitating earlier infalls for the satellites in order to match their measured present-day \HI masses. Both alternate mass-loading factors require infall times that are less likely for our sample than the fiducial \cite{Christensen2016} mass-loading factors (see Figure \ref{figure:antlia_infalls}). \par

\bsp	
\label{lastpage}
\end{document}